\documentclass[letterpaper,american,prl,twocolumn,amsmath,amssymb,showpacs,superscriptaddress]{revtex4-1}
\usepackage[T1]{fontenc}
\usepackage{babel}
\usepackage{prettyref}
\usepackage{amsbsy}
\usepackage{amstext}
\usepackage{graphicx}
\usepackage[unicode=true,
 bookmarks=true,bookmarksnumbered=false,bookmarksopen=false,
 breaklinks=false,pdfborder={0 0 1},backref=false,colorlinks=false]
 {hyperref}
\hypersetup{pdftitle={Noise dynamically suppresses chaos in random neural networks},
 pdfauthor={Goedeke, Schuecker, Helias},
 pdfsubject={Manuscript for PRL},
 pdfkeywords={Random neural networks, chaos},
 colorlinks=true,linkcolor=black,citecolor=black,urlcolor=black,filecolor=black}
\usepackage{breakurl}

\makeatletter

\newrefformat{cap}{\hyperref[#1]{Figure~\ref{#1}}}
\newrefformat{fig}{\hyperref[#1]{Figure~\ref{#1}}}
\newrefformat{tab}{\hyperref[#1]{Table ~\ref{#1}}}
\newrefformat{sec}{\hyperref[#1]{Section~\ref{#1}}}
\newrefformat{sub}{\hyperref[#1]{Section~\ref{#1}}}
\newrefformat{cha}{\hyperref[#1]{Chapter~\ref{#1}}}
\newrefformat{app}{\hyperref[#1]{Appendix~\ref{#1}}}
\usepackage{dsfont}

\makeatother

\begin{document}
\global\long\def\erf{\mathrm{erf}}
\global\long\def\LN{\mathcal{L}_{0}}
\global\long\def\LO{\mathcal{L}_{1}}
\global\long\def\mV{\:\mathrm{mV}}
\global\long\def\ms{\:\mathrm{ms}}
\global\long\def\Hz{\:\mathrm{Hz}}
\global\long\def\D{\mathcal{D}}
\global\long\def\J{\mathbf{J}}
\global\long\def\one{\mathbf{1}}
\global\long\def\e{\mathbf{e}}
\global\long\def\Cpp{\mathcal{K}_{\phi^{\prime}\phi^{\prime}}^{(0)}}
\global\long\def\CCpp{C_{\phi^{\prime}\phi^{\prime}}^{(0)}}
\global\long\def\Cppj{\mathcal{K}_{\phi_{j}^{\prime}\phi_{j}^{\prime}}^{(0)}}
\global\long\def\tx{\tilde{x}}
\global\long\def\xo{x^{(0)}}
\global\long\def\xii{x^{(1)}}
\global\long\def\txi{\tilde{x}^{(1)}}
\global\long\def\bx{\mathbf{x}}
\global\long\def\tbx{\tilde{\mathbf{x}}}
\global\long\def\bl{\mathbf{l}}
\global\long\def\bh{\mathbf{h}}
\global\long\def\bJ{\mathbf{J}}
\global\long\def\bN{\mathcal{N}}
\global\long\def\bH{\mathbf{H}}
\global\long\def\bK{\mathbf{K}}
\global\long\def\bxo{\bx^{(0)}}
\global\long\def\tbxo{\tilde{\bx}^{(0)}}
\global\long\def\bxi{\bx^{(1)}}
\global\long\def\tbxi{\tilde{\bx}^{(1)}}
\global\long\def\tbxi{\tbx^{(1)}}
\global\long\def\tpsi{\tilde{\psi}}
\global\long\def\Cxi{C_{x^{(1)}x^{(1)}}}
\global\long\def\bxxi{\mathbf{\xi}}
\global\long\def\N{\mathcal{N}}
\global\long\def\bW{\mathbf{W}}
\global\long\def\bon{\mathbf{1}}
\global\long\def\unity{\mathds{1}}

\global\long\def\D{\mathcal{D}}
\global\long\def\J{\mathbf{J}}
\global\long\def\one{\mathbf{1}}
\global\long\def\e{\mathbf{e}}
\global\long\def\Cpp{\mathcal{K}_{\phi^{\prime}\phi^{\prime}}^{(0)}}
\global\long\def\CCpp{C_{\phi^{\prime}\phi^{\prime}}^{(0)}}
\global\long\def\Cppj{\mathcal{K}_{\phi_{j}^{\prime}\phi_{j}^{\prime}}^{(0)}}
\global\long\def\tx{\tilde{x}}
\global\long\def\xo{x^{(0)}}
\global\long\def\xii{x^{(1)}}
\global\long\def\txi{\tilde{x}^{(1)}}
\global\long\def\bx{\mathbf{x}}
\global\long\def\tbx{\tilde{\mathbf{x}}}
\global\long\def\bl{\mathbf{j}}
\global\long\def\tbj{\tilde{\mathbf{j}}}
\global\long\def\bk{\mathbf{k}}
\global\long\def\tbk{\tilde{\mathbf{k}}}
\global\long\def\bh{\mathbf{h}}
\global\long\def\bJ{\mathbf{J}}
\global\long\def\bN{\mathcal{N}}
\global\long\def\bH{\mathbf{H}}
\global\long\def\bK{\mathbf{K}}
\global\long\def\bxo{\bx^{(0)}}
\global\long\def\tbxo{\tilde{\bx}^{(0)}}
\global\long\def\bxi{\bx^{(1)}}
\global\long\def\tbxi{\tilde{\bx}^{(1)}}
\global\long\def\tbxi{\tbx^{(1)}}
\global\long\def\tpsi{\tilde{\psi}}
\global\long\def\Cxi{C_{x^{(1)}x^{(1)}}}
\global\long\def\bxxi{\mathbf{\xi}}
\global\long\def\N{\mathcal{N}}
\global\long\def\bW{\mathbf{W}}
\global\long\def\bon{\mathbf{1}}
\global\long\def\tj{\tilde{j}}
\global\long\def\tJ{\tilde{J}}
\global\long\def\Z{\mathcal{Z}}
\global\long\def\SOM{S_{\mathrm{OM}}}
\global\long\def\SMSR{S_{\mathrm{MSR}}}

\global\long\def\cX{\mathcal{X}}
\global\long\def\cK{\mathcal{K}}
\global\long\def\cF{\mathcal{F}}
\global\long\def\unity{\mathds{1}}

\global\long\def\D{\mathcal{D}}
\global\long\def\bx{\mathbf{x}}
\global\long\def\bl{\mathbf{l}}
\global\long\def\bh{\mathbf{h}}
\global\long\def\bJ{\mathbf{J}}
\global\long\def\N{\mathcal{N}}
\global\long\def\hh{\hat{h}}
\global\long\def\bhh{\mathbf{\hh}}
\global\long\def\T{\mathrm{T}}
\global\long\def\by{\mathrm{\mathbf{y}}}
\global\long\def\diag{\mathrm{diag}}
\global\long\def\Ftr#1#2{\mathcal{F}\left[#1\right]\left(#2\right)}
\global\long\def\iFtr#1#2{\mathfrak{\mathcal{F}^{-1}}\left[#1\right]\left(#2\right)}
\global\long\def\D{\mathcal{D}}
\global\long\def\T{\mathrm{T}}
\global\long\def\Gammafl{\Gamma_{\mathrm{fl}}}
\global\long\def\gammafl{\gamma_{\mathrm{fl}}}
\global\long\def\E#1{\left\langle #1\right\rangle }

\global\long\def\M{\mathrm{M}}

\title{Optimal sequence memory in driven random networks}

\author{Jannis Schuecker$^{*}$ }

\affiliation{Institute of Neuroscience and Medicine (INM-6) and Institute for
Advanced Simulation (IAS-6) and JARA BRAIN Institute I, Jülich Research
Centre, Jülich, Germany}

\author{Sven Goedeke$^{*}$ }

\affiliation{Institute of Neuroscience and Medicine (INM-6) and Institute for
Advanced Simulation (IAS-6) and JARA BRAIN Institute I, Jülich Research
Centre, Jülich, Germany}

\author{Moritz Helias}

\affiliation{Institute of Neuroscience and Medicine (INM-6) and Institute for
Advanced Simulation (IAS-6) and JARA BRAIN Institute I, Jülich Research
Centre, Jülich, Germany}

\affiliation{Department of Physics, Faculty 1, RWTH Aachen University, Aachen,
Germany}

\date{\today}

\begin{abstract}
Autonomous randomly coupled neural networks display a transition to
chaos at a critical coupling strength. We here investigate the effect
of a time-varying input on the onset of chaos and the resulting consequences
for information processing. Dynamic mean-field theory yields the statistics
of the activity, the maximum Lyapunov exponent, and the memory capacity
of the network. We find an exact condition that determines the transition
from stable to chaotic dynamics and the sequential memory capacity
in closed form. The input suppresses chaos by a dynamic mechanism,
shifting the transition to significantly larger coupling strengths
than predicted by local stability analysis. Beyond linear stability,
a regime of coexistent locally expansive, but non-chaotic dynamics
emerges that optimizes the capacity of the network to store sequential
input.

$^{*}$ These authors contributed equally
\end{abstract}
\maketitle
Large random networks of neuron-like units can exhibit collective
chaotic dynamics \citep{Sompolinsky88_259,Vreeswijk96,Monteforte10_268104,Lajoie13_052901}.
Their information processing capabilities have been a focus in neuroscience
\citep{Maass02_2531} and in machine learning \citep{Jaeger04_87}
and show optimal performance close to the transition to chaos \citep{Legenstein07_323,Sussillo09_544,Toyoizumi11_051908}.
Due to its rich chaotic dynamics, the seminal network model by Sompolinsky
et al. \citep{Sompolinsky88_259} until today serves as a model for
various activity patterns observed in working memory tasks \citep{Barak13_214,Rajan16_128,Li16_459,Murray16_201619449},
motor control \citep{Laje13_925}, and perceptual decision making
\citep{Mante13_78}. The interplay between a time-dependent input
signal and the dynamical state of the network, however, is poorly
understood; notwithstanding consequences for information processing. 

In the absence of a signal the network dynamics is autonomous. Networks
of randomly coupled rate neurons display a transition from a fixed
point to chaotic fluctuations at a critical coupling strength \citep{Sompolinsky88_259},
illustrated in \prettyref{fig:potential_autocorrelation}a. The transition
is well understood by dynamic mean-field theory, originally developed
for spin glasses \citep{Sompolinsky81,Sompolinsky82_6860}. The onset
of chaos is equivalent to the emergence of a non-zero, decaying autocorrelation
function, whose decay time diverges at the transition. This equivalence
has been used in several subsequent studies \citep{rajan10_011903,Aljadeff15_088101,Harish15_e1004266}.
Furthermore a tight relationship to random matrix theory exists: the
transition happens precisely when the fixed point becomes linearly
unstable, which identifies the spectral radius of the random connectivity
matrix \citep{Sommers88,Rajan06} as the parameter controlling the
transition. 

These relations, however, lose their validity in the presence of
fluctuating input: stochasticity per se decorrelates the network activity
even if the dynamics is stable (\prettyref{fig:potential_autocorrelation}b),
so that a decaying autocorrelation function does not necessarily indicate
chaos. The stochastic drive, furthermore, causes perpetual fluctuations
also in the regular regime. Therefore, a transition to chaos, if existent
at all, must be of qualitatively different kind than the transition
from the silent fixed point in the autonomous case. Time-dependent
driving has indeed been found to stabilize network dynamics \citep{molgedey92_3717,rajan10_011903}.
However, the mechanism is only understood for low-dimensional systems
in the context of chaos synchronization by noise \citep{Zhou02_230602},
in networks driven by deterministic signals \citep{rajan10_011903},
and in systems with time-discrete dynamics \citep{molgedey92_3717}.
In the latter model, the effect of the fluctuating input on the transition
to chaos is completely captured by its influence on the spectral radius
of the Jacobian. Its single neuron dynamics, moreover, does not possess
non-trivial temporal correlations. But these temporal correlations
are indeed essential for the transition to chaos and for information
processing in time-continuous systems, as we will show here.

Realistic continuous-time network models can generate complex but
controlled responses to input \citep{Sussillo09_544} that resemble
activity patterns observed in motor cortex. In particular, the dynamical
state of the network plays a crucial role during the involved learning
process. However, the effect of the input on the dynamical state has
remained obscure.

To investigate the generic influence of external input on the network
state, we include additive white noise in the seminal model by \citet{Sompolinsky88_259}
and develop the dynamic mean-field theory for the resulting stochastic
continuous-time dynamics. In contrast to the original work, we here
reformulate the problem in terms of the functional formalism for stochastic
differential equations \citep{Martin73,janssen1976_377,dedominicis1976_247,DeDomincis78_353,Schuecker16b_arxiv}.
The application of the auxiliary field formulation known from large
$N$ field theory \citep{Moshe03} then allows us to derive the mean-field
equations by a saddle point approximation. We find that the autocorrelation
function is formally identical to the motion of a classical particle
in a potential, where the noise amounts to an initial kinetic energy.
We then determine the maximum Lyapunov exponent \citep{Eckmann85}
by considering two copies of the system with different initial conditions
\citep{Derrida86_45} in a replica calculation. Our main result is
a closed-form condition for the transition from stable to chaotic
dynamics. We find that the input suppresses chaos significantly more
strongly than expected from time-local linear stability, the criterion
valid in time-discrete systems. This observation is explained by a
dynamic effect: the decrease of the maximum Lyapunov exponent is related
to the sharpening of the autocorrelation function by the fluctuating
drive. The regime in the phase diagram between local instability,
as indicated by the spectral radius of the Jacobian, and transition
to chaos, corresponding to a positive maximum Lypunov exponent, constitutes
an as yet unreported dynamical regime that combines locally expansive
dynamics with asymptotic stability. Moreover, in contrast to the autonomous
case, the decay time of the autocorrelation function does not diverge
at the transition. Its peak is strongly reduced by the input and occurs
slightly above the critical coupling strength.

To study information processing capabilities we evaluate the capacity
to reconstruct a past input signal by a linear readout of the present
state, the so-called memory curve \citep{Jaeger01_memory}. Dynamic
mean-field theory and a replica calculation lead to a closed form
expression for the memory curve. We find that the memory capacity
peaks within the expansive, non-chaotic regime, indicating that locally
expansive while asymptotically stable dynamics is beneficial to store
input sequences in the dynamics of the neural network.

\section{Dynamic Mean-field equation}

We study the continuous-time dynamics of a random network of $N$
neurons, whose states $x_{i}(t)\in\mathbb{R},\ i=1,\dots,N,$ evolve
according to the system of stochastic differential equations
\begin{align}
\frac{dx_{i}}{dt} & =-x_{i}+\sum_{j=1}^{N}J_{ij}\phi(x_{j})+\xi_{i}(t)\,.\label{eq:diffeq_motion}
\end{align}
The $J_{ij}$ are independent and identically Gaussian distributed
random coupling weights with zero mean and variance $g^{2}/N$, where
the intensive gain parameter $g$ controls the recurrent coupling
strength or, equivalently, the weight heterogeneity of the network.
We further exclude self-coupling, setting $J_{ii}=0$. The time-varying
inputs $\xi_{i}(t)$ are pairwise independent Gaussian white-noise
processes with autocorrelation function $\langle\xi_{i}(t)\xi_{j}(s)\rangle=2\sigma^{2}\delta_{ij}\delta(t-s)$.
We choose the sigmoidal transfer function $\phi(x)=\tanh(x)$, so
that without input, for $\sigma=0$, the model agrees with the autonomous
one studied in \citep{Sompolinsky88_259}.

The dynamical system \eqref{eq:diffeq_motion} contains two sources
of randomness: the quenched disorder due to the random coupling weights
and temporally fluctuating drive. A particular realization of the
random couplings $J_{ij}$ defines a fixed network configuration and
its dynamical properties usually vary between different realizations.
For large network size $N$, however, certain quantities are self-averaging,
meaning that their values for a typical realization can be obtained
by an average over network configurations \citep{Fischer91}. An important
example is the population-averaged autocorrelation function.

We here derive a dynamic mean-field theory that describes the statistical
properties of the system under the joint distribution of disorder,
noise, and possibly random initial conditions in the limit of large
network size $N\rightarrow\infty$. The theory can be derived via
a heuristic ``local chaos'' assumption \citep{Amari72_643} or using
a generating functional formulation \citep{Sompolinsky82_6860,Crisanti87_4922}.
We here follow the latter approach, because it casts the problem into
the established language of statistical field theory for which a wealth
of approximation techniques is available \citep{ZinnJustin96}. A
mathematically rigorous proof uses large deviation techniques \citep{Cabana13_211}.
The general idea is that for large network size $N$ the local recurrent
input $\sum_{j=1}^{N}J_{ij}\phi(x_{j})$ in \prettyref{eq:diffeq_motion}
approaches a Gaussian process with self-consistently determined statistics. 

We interpret the stochastic differential equations in the Ito-convention
\citep{Gardiner85} and formulate the problem \prettyref{eq:diffeq_motion}
in terms of a moment-generating functional $Z$. Using the Martin-Siggia-Rose-De
Dominicis-Janssen path integral formalism \citep{Martin73,DeDomincis78_353,Altland01}
we obtain 
\begin{align}
Z[\bl](\bJ) & =\int\D\bx\int\D\tbx\,\exp\Big(S_{0}[\bx,\tbx]-\tbx^{\T}\bJ\phi\left(\bx\right)+\bl^{\T}\bx\Big)\label{eq:Z_j}\\
\text{with }S_{0}[\bx,\tbx] & =\tbx^{\T}\left(\partial_{t}+1\right)\bx+\sigma^{2}\tbx^{\T}\tbx,\label{eq:def_S0}
\end{align}
where $\bx^{\T}\by=\sum_{i}\int\,x_{i}(t)y_{i}(t)\,dt$ denotes the
scalar product in time and in neuron space and $\tbx$ and $\bl$
represent a response field and a source field, respectively. The measures
are defined as $\int\D\bx=\lim_{M\to\infty}\Pi_{k=1}^{N}\Pi_{l=1}^{M}\int_{-\infty}^{\infty}dx_{k}^{l}$
and $\int\D\tbx=\lim_{M\to\infty}\Pi_{k=1}^{N}\Pi_{l=1}^{M}\int_{-i\infty}^{i\infty}(2\pi i)^{-1}d\tilde{x}_{k}^{l}$
with the subscript $k$ denoting the $k$-th unit and the superscript
$l$ denoting the $l$-th time slice. The action $S_{0}$ in \eqref{eq:def_S0}
contains all single unit properties, therefore excluding the coupling
term $-\tbx^{\T}\bJ\phi\left(\bx\right)$, which is written explicitly
in \prettyref{eq:Z_j}.

Assuming that the dynamics is self-averaging, we average over the
quenched disorder in the connectivity $\bJ$ and perform a saddle-point
approximation (\prettyref{app:Derivation-of-mean-field}). The resulting
functional factorizes into $N$ terms
\begin{align}
\bar{Z}^{\ast} & \propto\int\D x\int\D\tilde{x}\,\exp\,\Big(S_{0}[x,\tx]+\frac{g^{2}}{2}\tx^{\T}C_{\phi(x)\phi(x)}\tx\Big),\label{eq:Z_bar_star}
\end{align}
with $C_{\phi(x)\phi(x)}(t,s):=\langle\phi(x(t))\phi(x(s))\rangle$
denoting the average autocorrelation function of the non-linearly
transformed activity of the units \prettyref{eq:saddle_Q1_Q2} and
$\tx^{\T}C_{\phi(x)\phi(x)}\tx:=\iint\,dt\,ds\,\tx(t)\,C_{\phi(x)\phi(x)}(t,s)\,\tx(s)$.
The factorization reduces the network to $N$ non-interacting units,
each on a background of an independent Gaussian noise with identical
self-consistently determined statistics. At this level of approximation,
the problem is hence equivalent to a single unit system. The effective
equation of motion corresponding to this system can be read from \prettyref{eq:Z_bar_star}
(\prettyref{app:Derivation-of-mean-field}) 
\begin{eqnarray}
\frac{dx}{dt} & = & -x+\eta(t)+\xi(t).\label{eq:mean-field_diffeq}
\end{eqnarray}

Here, $\xi(t)$ is a Gaussian white-noise process as in \prettyref{eq:diffeq_motion},
independent of $\eta(t)$. The centralized Gaussian process $\eta(t)$
is fully specified by its autocorrelation function
\begin{align}
\langle\eta(t)\eta(s)\rangle & =g^{2}C_{\phi(x)\phi(x)}(t,s).\label{eq:mean-field_inputcorrfun}
\end{align}

\section{Effective equation of motion of the autocorrelation}

\begin{figure}
\includegraphics{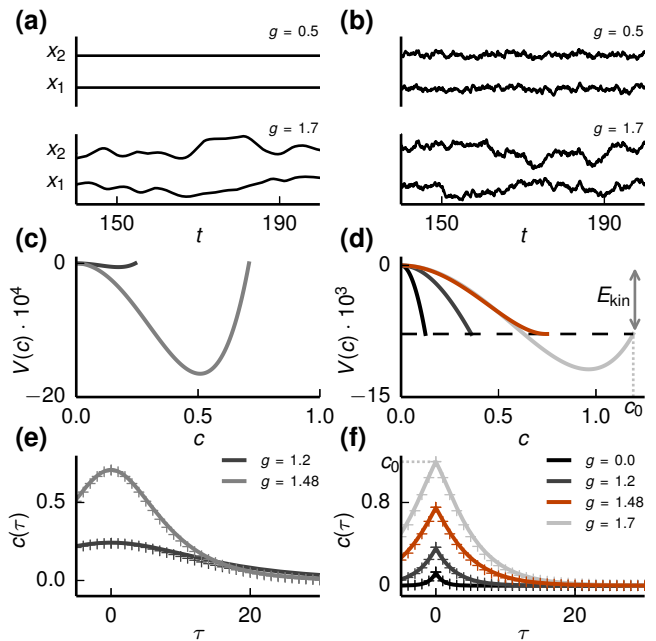}\caption{\textbf{\label{fig:potential_autocorrelation}Activity statistics
of autonomous and driven network. }Autonomous case $\sigma=0$ (left
column) and driven case $\sigma=\sqrt{0.125}$ (right column). \textbf{Upper
row}: Simulated trajectories of two example neurons for sub-critical
$g=0.5$ (upper part of vertical axis) and super-critical coupling
$g=1.7$ (lower part of vertical axis). \textbf{Middle row}: Classical
potential \prettyref{eq:classic_potential} with self-consistently
determined variance $c_{0}$ following from energy conservation \prettyref{eq:self_consistent_c0}
for different coupling strengths $g$ (corresponding legends in lower
row); dashed horizontal line at minus initial kinetic energy $E_{\mathrm{kin}}=\sigma^{4}/2$.
In the driven case the critical coupling $g_{c}=1.48$ from eq. \prettyref{eq:transition_crit}
is shown in red.\textbf{ Lower row}: Self-consistent autocorrelation
function (solid line) compared to simulations (crosses). The variance
(peak height) $c_{0}$ corresponds to the largest value of $c$ at
which the potential (middle row) is defined, indicated for $g=1.7$
with gray dotted lines in (d) and (f). Network size in simulations
is $N=10000$.}
\end{figure}

Our goal is to determine the mean-field autocorrelation function $\langle x(t)x(s)\rangle$,
which, by self-averaging, also describes the population-averaged autocorrelation
function. Assuming that $x(t)$ is a stationary process, $c(\tau)=\langle x(t+\tau)x(t)\rangle$
obeys the differential equation (\prettyref{app:Stationary-process})
\begin{align}
\ddot{c}=\frac{d^{2}c}{d\tau^{2}} & =c-g^{2}f_{\phi}(c,c_{0})-2\sigma^{2}\delta(\tau)\label{eq:diffeq_c}
\end{align}
with $c_{0}=c(0)$. The Dirac-$\delta$ inhomogeneity originates from
the white-noise autocorrelation function of the time-varying input
and is absent in \citep{Sompolinsky88_259}. The same inhomogeneity
arises from Poisson spiking noise with $2\sigma^{2}=g^{2}r$ \citep{Kadmon15_041030},
where $r$ is the population-averaged firing rate. In \prettyref{eq:diffeq_c}
we write $f_{\phi}(c(\tau),c_{0})=C_{\phi(x)\phi(x)}(t+\tau,t)$,
introducing the notation
\begin{align}
f_{u}(c,c_{0})\label{eq:def_f}\\
= & \iint\,u\left(\sqrt{c_{0}-\tfrac{c^{2}}{c_{0}}}\,z_{1}+\tfrac{c}{\sqrt{c_{0}}}\,z_{2}\right)\,u\left(\sqrt{c_{0}}\,z_{2}\right)\,Dz_{1}Dz_{2}\nonumber 
\end{align}
for an arbitrary function $u(x)$ and the Gaussian integration measures
$Dz_{i}=\exp(-z_{i}^{2}/2)/\sqrt{2\pi}\,dz_{i},\quad i=1,2$. This
representation holds since $x(t)$ is itself a Gaussian process. Note
that \prettyref{eq:def_f} reduces to a one-dimensional integral for
$f_{u}(c_{0},c_{0})=\langle u(\sqrt{c_{0}}z_{1})^{2}\rangle$ and
$f_{u}(0,c_{0})=\langle u(\sqrt{c_{0}}z_{1})\rangle^{2}$.

We formulate \prettyref{eq:diffeq_c} as the one-dimensional motion
of a classical particle in a potential: 
\begin{align}
\ddot{c} & =-V^{\prime}(c)-2\sigma^{2}\delta(\tau)\,,\label{eq:equation_of_motion}
\end{align}
where we define 
\begin{align}
V(c)=V(c;c_{0})= & -\frac{1}{2}c^{2}+g^{2}f_{\Phi}(c,c_{0})-g^{2}f_{\Phi}(0,c_{0})\,,\label{eq:classic_potential}
\end{align}
with $\Phi(x)=\int_{0}^{x}\phi(y)\,dy$ and $\partial/\partial c\,f_{\Phi}(c,c_{0})=f_{\phi}(c,c_{0})$
following from Price's theorem \citep{Price58_69,Papoulis91}. The
autocorrelation $c(\tau)$ here plays the role of the position of
the particle and the time lag $\tau$ the role of time. The potential
\prettyref{eq:classic_potential} depends on the initial value $c_{0}$,
which has to be determined self-consistently. We obtain $c_{0}$ from
classical energy conservation $\dot{c}^{2}/2+V(c)=\text{constant}$.
Considering $\tau\ge0$ and the symmetry of $c(\tau)$, the fluctuating
drive in \prettyref{eq:equation_of_motion} amounts to an initial
velocity $\dot{c}(0+)=-\sigma^{2}$ and thus to the kinetic energy
$\dot{c}^{2}(0+)/2=\sigma^{4}/2$. Since $|c(\tau)|\le c_{0}$, the
solution $c(\tau)$ and its first derivative must approach zero as
$\tau\rightarrow\infty$. Thus we obtain the self-consistency condition
for $c_{0}$ as
\begin{equation}
\frac{1}{2}\sigma^{4}+V(c_{0};c_{0})=V(0;c_{0})=0.\label{eq:self_consistent_c0}
\end{equation}
For the autonomous case, \prettyref{fig:potential_autocorrelation}c,e
shows the resulting potential and the corresponding self-consistent
autocorrelation function $c(\tau)$ in the chaotic regime. Approaching
the transition from above, $g\rightarrow g_{c}=1,$ the amplitude
$c_{0}$ vanishes and the decay time of $c(\tau)$ diverges \citep{Sompolinsky88_259}.
This picture breaks down in the driven case (\prettyref{fig:potential_autocorrelation}d,f),
where $c_{0}$ is always nonzero, $c(\tau)$ decays with finite time
scale and has a kink at zero. The mean-field prediction is in excellent
agreement with the population-averaged autocorrelation function obtained
from numerical simulations of one network instance showing that the
self-averaging property is fulfilled. In the following we derive a
condition for the transition from stable to chaotic dynamics in the
presence of the time-varying input.

\section{Effect of Input on the transition to chaos}

The maximum Lyapunov exponent quantifies how sensitive the dynamics
depends on the initial conditions \citep{Eckmann85}. It measures
the asymptotic growth rate of infinitesimal perturbations. For stochastic
dynamics the stability of the solution for a fixed realization of
the noise or equivalently the stochastic input is also characterized
by the maximum Lyapunov exponent \footnote{The theory of random dynamical systems makes this more precise; a
brief overview is given in \citep{Lajoie13_052901}.}: If it is negative, trajectories with different initial conditions
converge to the same time-dependent solution; the dynamics is stable.
If it is positive, the distance between two initially arbitrary close
trajectories grows exponentially in time; the dynamics exhibits sensitive
dependence on initial conditions and is hence chaotic.

We derive the maximum Lyapunov exponent by using dynamic mean-field
theory. To this end, we consider two copies of the network distinguished
by superscripts $\alpha\in\{1,2\}$. These copies, or replicas, have
identical coupling matrix $\bJ$ and, for $\sigma>0$, are subject
to the same realization of the stochastic input $\xi_{i}(t)$. The
maximum Lyapunov exponent can be defined as the asymptotic growth
rate of the Euclidean distance between trajectories of the two copies:
\begin{align*}
\lambda_{\mathrm{max}} & =\lim_{t\rightarrow\infty}\lim_{||\bx^{1}(0)-\bx^{2}(0)||\rightarrow0}\frac{1}{2t}\ln\left(\frac{||\bx^{1}(t)-\bx^{2}(t)||^{2}}{||\bx^{1}(0)-\bx^{2}(0)||^{2}}\right)\,.
\end{align*}

We now follow an idea by \citet{Derrida86_45} and exploit the self-averaging
property of population-averaged correlation functions, i.e, $\frac{1}{N}\sum_{i=1}^{N}x_{i}^{\alpha}(t)x_{i}^{\beta}(s)\approx c^{\alpha\beta}(t,s)$,
where $c^{\alpha\beta}$ denote the correlation functions averaged
over the realization of the couplings. We express the mean squared
Euclidean distance as
\begin{align*}
\frac{1}{N}\sum_{i=1}^{N}\left(x_{i}^{1}(t)-x_{i}^{2}(t)\right)^{2} & \approx c^{11}(t,t)+c^{22}(t,t)-2c^{12}(t,t)\,\\
 & \equiv d(t)\,,
\end{align*}
where we defined the mean-field squared distance $d(t)$. Thus the
asymptotic growth rate of $d(t)$ provides us with a mean-field description
of the maximum Lyapunov exponent. To obtain this growth rate we first
consider
\begin{align}
d(t,s) & =c^{11}(t,s)+c^{22}(t,s)-c^{12}(t,s)-c^{21}(t,s)\label{eq:mean-squared-distance}
\end{align}
with the obvious property $d(t)=d(t,t)$. We then determine the temporal
evolution of $d(t,s)$ for infinitesimally perturbed initial conditions
$||\bx^{1}(0)-\bx^{2}(0)||=\epsilon$. To this end it is again convenient
to use a generating functional that captures the joint statistics
of the two systems and in addition allows averaging over the quenched
disorder \citep[see also ][Appendix 23, last remark]{ZinnJustin96}.
The generating functional describing the two copies is defined analogously
to the single system \prettyref{eq:Z_j} as\begin{widetext}

\begin{align}
Z[\{\bl^{\alpha}\}_{\alpha\in\{1,2\}}](\bJ) & =\Pi_{\alpha=1}^{2}\Big\{\int\D\bx^{\alpha}\int\D\tbx^{\alpha}\,\exp\Big(S_{0}[\bx^{\alpha},\tbx^{\alpha}]-\tbx^{\alpha\T}\bJ\phi\left(\bx^{\alpha}\right)+\bl^{\alpha\T}\bx^{\alpha}\Big)\Big\}\exp\left(2\sigma^{2}\tbx^{1\T}\tbx^{2}\right)\label{eq:Z_pair}
\end{align}
\end{widetext}with the single system ``free action'' $S_{0}[\bx,\tbx]$
defined in \prettyref{eq:def_S0}. The factor in the last line results
from the identical external input in the two copies and effectively
couples the two systems. We also note that the coupling matrix $\bJ$
is the same in both copies.

Averaging \prettyref{eq:Z_pair} over the quenched disorder of the
random coupling matrix $\bJ$ and performing a saddle-point approximation
we obtain a pair of effective dynamical equations (\prettyref{app:pair_of_systems}),
\begin{align}
\left(\partial_{t}+1\right)x^{\alpha}(t) & =\xi(t)+\eta^{\alpha}(t)\,,\quad\alpha\in\{1,2\}\,,\label{eq:effective_pair_eq}
\end{align}
together with a set of self-consistency equations for the statistics
of the noises $\eta^{\alpha}$
\begin{align}
\langle\eta^{\alpha}(t)\,\eta^{\beta}(s)\rangle & =g^{2}\,\langle\phi(x^{\alpha}(t))\phi(x^{\beta}(s))\rangle.\label{eq:Leffective_pair_noise}
\end{align}
Now, there are two terms which introduce correlations between the
two copies. First the common temporal fluctuations $\xi(t)$ injected
into both systems. Second the effective noises $\eta^{\alpha}$ and
$\eta^{\beta}$ are correlated between replicas \prettyref{eq:Leffective_pair_noise},
arising from the two systems having the same coupling $\bJ$ in each
realization. The origin of the latter coupling is hence of static
nature.

\begin{figure}
\includegraphics{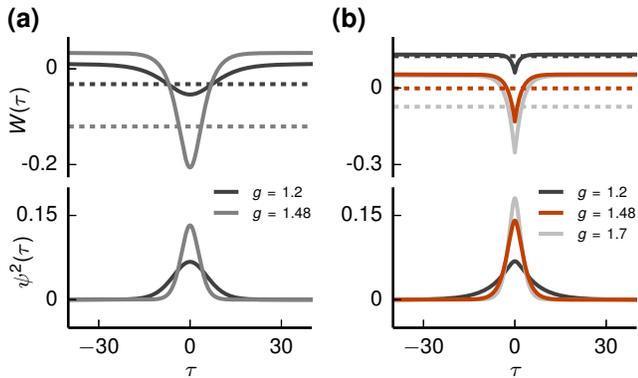}\caption{\textbf{Ground state of Schrödinger equation determines Lyapunov exponent.
}Upper part of vertical axis: Quantum potential $W$ (solid curve)
and ground state energy $E_{0}$ (dashed line) for autonomous case
\textbf{(a)} and driven case \textbf{(b)} for $\sigma=\sqrt{0.125}$.
Lower part of vertical axis: Corresponding squared ground state wave
function. Parameters as in \prettyref{fig:potential_autocorrelation}
(driven case for $g=0$ left out).\label{fig:Solution-of-Schroedinger} }
\end{figure}

The distance \prettyref{eq:mean-squared-distance} between the two
copies is given by the auto-correlations of the single systems and
the cross-correlations between them. We consider the case where both
copies are prepared with identical initial conditions and thus are
fully synchronized: the cross-correlation $c^{12}$ initially equals
the auto-correlations $c^{11}$, $c^{22}$. The latter are identical
to the single-system autocorrelation function $c$, because the marginal
statistics of each subsystem is not affected by the mere presence
of the respective other system.  An increase of the distance $d(t)$,
by \prettyref{eq:mean-squared-distance}, amounts to a decline of
$c^{12}$ away from its initial value $c$. Here $c$ is the stationary
autocorrelation as we are interested in the Lyapunov exponent averaged
over initial conditions drawn from the stationary distribution. To
determine the growth rate in the limit of small distances $d(0)\propto\epsilon$
between the two copies we therefore expand the cross-correlation around
its stationary solution $c^{12}(t,s)=c(t-s)+\epsilon\,k^{(1)}(t,s)\,,\:\epsilon\ll1$,
which leads to an equation of motion for the first order deflection
(\prettyref{app:Derivation-of-variational})

\begin{equation}
\left(\partial_{t}+1\right)\left(\partial_{s}+1\right)k^{(1)}(t,s)=g^{2}f_{\phi'}(c(t-s),c_{0})\,k^{(1)}(t,s)\label{eq:variational_equation}
\end{equation}
with $d(t)=-2\epsilon\,k^{(1)}(t,t)$.

A separation ansatz in the coordinates $\tau=t-s$ and $T=t+s$ then
yields an eigenvalue problem in the form of a time-independent Schrödinger
equation \citep{Sompolinsky88_259,Kadmon15_041030} (\prettyref{app:Schroedinger-equation})
\begin{equation}
\left[-\partial_{\tau}^{2}+W(\tau)\right]\,\psi(\tau)=E\,\psi(\tau),\label{eq:Schroedinger}
\end{equation}
where now $\tau$ plays the role of a spatial coordinate. Here, the
quantum potential $W(\tau)=-V^{\prime\prime}(c(\tau))=1-g^{2}f_{\phi^{\prime}}(c(\tau),c_{0})$
is given by the negative second derivative of the classical potential
$V(c)$ evaluated along the self-consistent autocorrelation function
$c(\tau)$. The ground state energy $E_{0}$ of \prettyref{eq:Schroedinger}
determines the asymptotic growth rate of $k^{(1)}(t,t)$ as $t\rightarrow\infty$
and, hence, the maximum Lyapunov exponent via $\lambda_{\mathrm{max}}=-1+\sqrt{1-E_{0}}$
\prettyref{eq:Lambda_max}. Therefore, the dynamics is predicted to
become chaotic if $E_{0}<0$. The quantum potential together with
the solution for the ground state energy and wave function is shown
in \prettyref{fig:Solution-of-Schroedinger}. The latter are obtained
as solutions of a finite difference discretization of \prettyref{eq:Schroedinger}.

\begin{figure}
\includegraphics{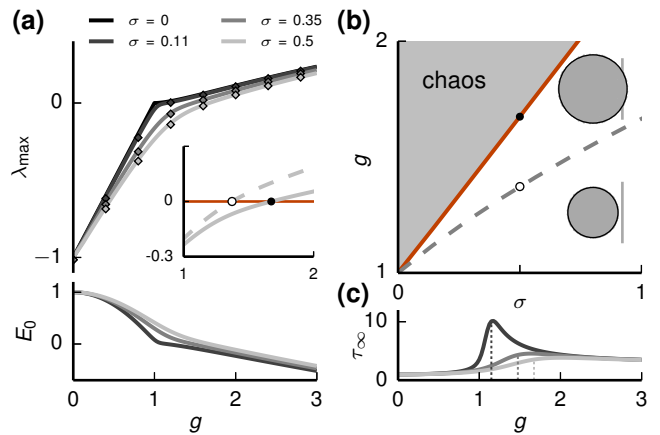}\caption{\textbf{Transition to chaos. (a)} Upper part of vertical axis: Maximum
Lyapunov exponent $\lambda_{\mathrm{max}}$ as a function of the coupling
strength $g$ for different input amplitude levels. Mean-field prediction
(solid curve) and simulation (diamonds). Comparison to the upper bound
$-1+g\sqrt{\langle[\phi'(x)]^{2}\rangle}$ (dashed) for $\sigma=0.5$
in inset. Zero crossings marked with dots. Lower part of vertical
axis: Ground state energy $E_{0}$ as a function of $g$.\textbf{
(b)} Phase diagram with transition curve (solid red curve) obtained
from \prettyref{eq:transition_crit} and necessary condition (\prettyref{eq:local_instability}
with equal sign, gray dashed curve). Dots correspond to zero crossings
in inset in (a). Disk of eigenvalues of the Jacobian matrix in \prettyref{eq:evolution_y}
for $\sigma=0.8$ and $g=1.25$ (lower) and $g=2.0$ (upper) centered
at $-1$ in the complex plane (gray). Radius $\rho=g\sqrt{\langle[\phi'(x)]^{2}\rangle}$
from random matrix theory (black). Vertical line at zero.\textbf{
(c)} Asymptotic decay time $\tau_{\infty}$ of autocorrelation function.
Vertical dashed lines mark the transition to chaos. Color code as
in (a). Network size of simulations $N=5000$.\label{fig:Transition-to-chaos}}
\end{figure}

In the autonomous case, a decaying autocorrelation function corresponds
to a positive maximum Lyapunov exponent \citep{Sompolinsky88_259}.
This follows from the observation that for $g>1$ the derivative of
the self-consistent autocorrelation function $\dot{c}(\tau)$ solves
the Schrödinger equation with $E=0$. But as $\dot{c}(\tau)$ is an
eigenfunction with a single node it cannot be the ground state, which
has zero nodes. The ground state energy, which is necessarily lower,
must therefore be negative, $E_{0}<0$. So the dynamics is chaotic
and $\lambda_{\mathrm{max}}$ crosses zero at $g=1$ (\prettyref{fig:Transition-to-chaos}a).

In the presence of fluctuating drive, the maximum Lyapunov exponent
becomes positive at a critical coupling strength $g_{c}>1$; with
increasing input amplitude the transition shifts to larger values
(\prettyref{fig:Transition-to-chaos}a). The mean-field prediction
$\lambda_{\mathrm{max}}=-1+\sqrt{1-E_{0}}$ shows excellent agreement
with the maximum Lyapunov exponent obtained in simulations using a
standard algorithm \citep{Eckmann85}. Since the ground state energy
$E_{0}$ must be larger than the minimum $W(0)=1-g^{2}\langle[\phi^{\prime}(x)]^{2}\rangle$
of the quantum potential, an upper bound for $\lambda_{\mathrm{max}}$
is provided by $-1+g\sqrt{\langle[\phi'(x)]^{2}\rangle}$ leading
to a necessary condition
\begin{eqnarray}
g\sqrt{\langle[\phi'(x)]^{2}\rangle} & \ge & 1\label{eq:local_instability}
\end{eqnarray}
for chaotic dynamics. However, close to the transition $\lambda_{\mathrm{max}}$
is clearly smaller than the upper bound, which is a good approximation
only for small $g$ (\prettyref{fig:Transition-to-chaos}a, inset):
the actual transition occurs at substantially larger coupling strengths.
In contrast, for memoryless discrete-time dynamics the necessary condition
found here is also sufficient for the transition to chaos \citep[eq. 13]{molgedey92_3717}. 

The local linear stability of the dynamical system \prettyref{eq:diffeq_motion}
is analyzed via the variational equation 
\begin{align}
\frac{d}{dt}y_{i}(t) & =-y_{i}(t)+\sum_{j=1}^{N}J_{ij}\phi'(x_{j}(t))\,y_{j}(t)\,,\label{eq:evolution_y}
\end{align}
$i=1,\ldots,N,$ describing the temporal evolution of an infinitesimal
deviation $y_{i}(t)$ about a reference trajectory $x_{i}(t)$. Interestingly,
$\rho=g\sqrt{\langle[\phi'(x)]^{2}\rangle}$ (cf. \prettyref{eq:local_instability})
is also the radius of the disk formed by the eigenvalues of the Jacobian
matrix in the variational equation \prettyref{eq:evolution_y} estimated
by random matrix theory \citep{Sommers88,Rajan06}. Therefore, the
dynamics is expected to become locally unstable if this radius exceeds
unity, as shown in the inset in \prettyref{fig:Transition-to-chaos}b
displaying $\rho$ and the eigenvalues at an arbitrary point in time.
But even for the case with $\rho>1$ the system is not necessarily
chaotic. Hence, contrary to the autonomous case \citep{Sommers88,Sompolinsky88_259},
the transition to chaos is not predicted by random matrix theory.

To derive an exact condition for the transition we determine a ground
state with vanishing energy $E_{0}=0$. As in the autonomous case,
$\dot{c}(\tau)$ solves \prettyref{eq:Schroedinger} for $E=0$, except
at $\tau=0$ where it exhibits a jump, because $c(\tau)$ has a kink
due to the input \prettyref{eq:diffeq_c}. By linearity $|\dot{c}(\tau)|$
is a continuous and symmetric solution with zero nodes. Therefore,
if its derivative is continuous as well, requiring $\ddot{c}(0+)=0$,
it constitutes the searched for ground state. This is in contrast
to the autonomous case, where $\dot{c}(\tau)$ corresponds to the
first excited state. Consequently, with \prettyref{eq:diffeq_c} we
find the condition for the transition
\begin{align}
g_{c}^{2}\,f_{\phi}(c_{0},c_{0})-c_{0} & =0\,,\label{eq:transition_crit}
\end{align}
in which $c_{0}$ is determined by the self-consistency condition
\prettyref{eq:self_consistent_c0} resulting in the transition curve
$(g_{c},\sigma_{c})$ in parameter space (\prettyref{fig:Transition-to-chaos}b).
This reveals the relationship between the onset of chaos, the statistics
of the random coupling matrix, and the input amplitude.

From \prettyref{eq:transition_crit} follows that the system becomes
chaotic precisely when the variance $c_{0}$ of a typical single unit
equals the variance of its recurrent input from the network $g_{c}^{2}\langle\phi^{2}\rangle$.
At the transition the classical self-consistent potential $V(c;c_{0})$
has a horizontal tangent at $c_{0}$, while in the chaotic regime
a minimum emerges (\prettyref{fig:potential_autocorrelation}d). This
implies with \prettyref{eq:diffeq_c} that the curvature $\ddot{c}(0+)$
of the autocorrelation function at zero changes sign from positive
to negative (\prettyref{fig:potential_autocorrelation}f). Close to
the transition a standard perturbative approach shows that $\lambda_{\mathrm{max}}$
is proportional to $g^{2}\langle\phi^{2}(x)\rangle-c_{0}$, indicating
a self-stabilizing effect: since both terms grow with $g$, the growth
of their difference is attenuated, explaining why $\lambda_{\mathrm{max}}(g)$
bends down as the transition is approached (\prettyref{fig:Transition-to-chaos}a). 

The condition \prettyref{eq:transition_crit} predicts the transition
at significantly larger coupling strengths compared to the necessary
condition (\prettyref{fig:Transition-to-chaos}b), which is explained
as follows. For continuous-time dynamics the effect of fluctuating
input is twofold: First, because the slope $\phi^{\prime}(x)$ is
maximal at the origin, fluctuations reduce the averaged squared slope
in $g^{2}\langle[\phi'(x)]^{2}\rangle$, thereby stabilizing the dynamics.
This is an essentially static effect as it can be fully attributed
to the increase of the variance $c_{0}$ caused by the additional
input; static heterogeneous inputs would have the same effect. Second,
the input sharpens the autocorrelation function (\prettyref{fig:potential_autocorrelation}e,f)
and hence the quantum potential (\prettyref{fig:Solution-of-Schroedinger}).
This shifts the ground-state energy to larger values, further decreasing
the maximum Lyapunov exponent. Because this effect depends on the
temporal correlations, the input suppresses chaos by a dynamic mechanism
yielding stable dynamics even in the presence of local linear instability.

To understand this dynamic mechanism we return to the variational
equation \prettyref{eq:evolution_y}: its fundamental solution can
be regarded as a product of short-time propagator matrices, where
each factor has the same stability properties with unstable directions
given by the local Jacobian matrix at the respective time. Even though
the fraction of eigenvalues with positive real part stays approximately
constant, the corresponding unstable directions vary in time. The
sharpening of the autocorrelation function suggests that fluctuating
input causes a faster variation, such that perturbations cannot grow
in the direction of unstable modes, but rather decay asymptotically.

In low-dimensional systems the suppression of chaos by external fluctuations
is understood: Noise forces the system to visit regions of the phase
space with locally contracting dynamics more frequently \citep{Zhou02_230602}
so that contraction dominates expansion, in total yielding stable
asymptotic behavior. This mechanism is similar to the static stabilization
effect described above, where fluctuating drive causes the system
to sample regions of the phase space with smaller eigenvalues of the
Jacobian. The self-averaging high-dimensional system, however, has
a constant spectral radius over time and hence the dynamics is either
locally contracting or locally expanding for all times. While the
previous effects are explained by local stability, the dynamic suppression
of chaos found here is a genuinely time-dependent mechanism, explained
by the time evolution of the Jacobian.

In the autonomous case the time-scale of fluctuations diverges at
the transition to chaos \citep{Sompolinsky88_259}. We here consider
the effect of the input on the asymptotic decay time $\tau_{\infty}=1/\sqrt{1-g^{2}\langle\phi^{\prime}(x)\rangle^{2}}$
of the autocorrelation function (\prettyref{fig:Transition-to-chaos}c).
For weak input amplitude, the decay time peaks at the transition,
corresponding to the diverging time scale in the autonomous case.
For larger input amplitudes, the peak is strongly reduced and the
maximum decay time is attained above the transition.

\section{Information processing capabilities}

We expect the expansive, non-chaotic regime to be beneficial for information
processing: The local instability of the network ensures sufficient
initial amplification of the impinging external signal. The asymptotic
stability is required for the driving signal not to be corrupted by
the unbounded amplification of small variations of the input; it is
hence necessary to ensure generalization. In the following we investigate
these ideas quantitatively by considering the sequential memory capacity
of the network.

We focus on the component $z(t)=\frac{1}{\sqrt{N}}\,\sum_{i=1}^{N}\xi_{i}(t)$
of the input that is received by all neurons with equal strength.
In other words, the total input to each neuron is decomposed into
the signal $z(t)$ and the remaining inputs $\xi_{i}(t)-z(t)$ which
act as noise. We then consider the dynamical short-term memory defined
as the capacity to reconstruct the input $z(t)\,$ from the state
at time $t+\tau$ using a linear readout, $\sum_{i=1}^{K}w_{i}x_{i}(t+\tau)$,
where $K\le N$ is the number of readout neurons. The reconstruction
capacity as a function of the delay time $\tau$ yields the memory
curve $m(\tau)=1-\epsilon(\tau)$ \citep{Jaeger01_memory}, where
$\epsilon(\tau)$ is the minimal relative mean-squared error between
readout and signal. Alternatively this measure quantifies the fidelity
by which a sequence of past inputs can be reconstructed from the current
network activity.

\begin{figure}
\includegraphics{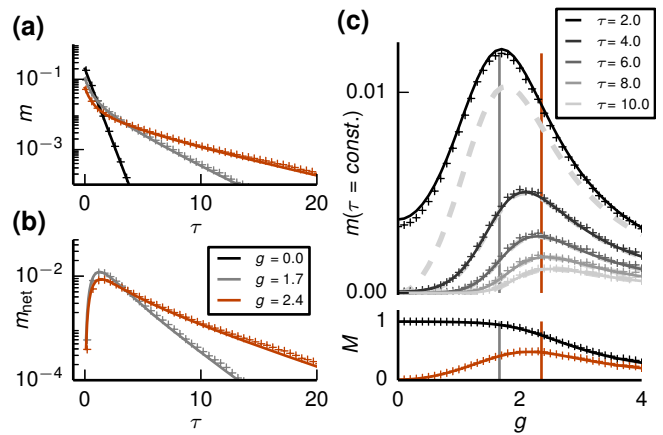}\caption{\textbf{Sequential memory.} Mean-field prediction (solid curves) and
simulation (crosses). \textbf{(a)} Memory \prettyref{eq:memory_curve}
as a function of time lag $\tau$ between signal and readout for different
coupling strengths encoded in color (legend in b).\textbf{ (b) }Network
contribution \prettyref{eq:memory_curve_net} to memory for different
coupling strengths $g$. \textbf{(c)} Upper part of vertical axis:
Memory at different time lags $\tau$ over coupling strength. Network
contribution to memory shown as dashed thick light-gray curves, which
coincide with total memory curves for $\tau\ge4$. Vertical gray line
marks local instability \prettyref{eq:local_instability} and vertical
red line marks transition to chaos (cf. \prettyref{fig:Transition-to-chaos}b).
Lower part of vertical axis: Memory capacity $M$ \prettyref{eq:memory_capacity}
(black) and network contribution to memory capacity $M_{\mathrm{net}}$
(red).\label{fig:Memory-capacity}}
\end{figure}

For optimal readout weights $\boldsymbol{w}$ that minimize $\epsilon(\tau)$,
the memory curve is given by \citep{Jaeger01_memory,Dambre12_514}

\begin{eqnarray}
m(\tau) & = & \frac{\langle\boldsymbol{x}(t+\tau)z(t)\rangle^{\mathrm{T}}\langle\boldsymbol{x}(t)\boldsymbol{x}(t)^{\mathrm{T}}\rangle^{-1}\langle\boldsymbol{x}(t+\tau)z(t)\rangle}{\langle z(t)^{2}\rangle}\,.\nonumber \\
\label{eq:memory_def}
\end{eqnarray}
We follow the approach by \citet{Toyoizumi11_051908} and neglect
the off-diagonal terms in $\langle\boldsymbol{x}\boldsymbol{x}^{\mathrm{T}}\rangle$,
which is justifiable for a sparse readout with $K\ll N$. Additionally,
for large $N$ the diagonal terms $\E{x_{i}^{2}}$ are given by their
mean-field value $c_{0}$, identical for all units. Determining the
memory curve \prettyref{eq:memory_def} then amounts to computing
the sum of squared correlation functions $\sum_{i}\langle x_{i}(t+\tau)z(t)\rangle^{2}$
between the signal and the network activity, which we obtain by a
replica calculation (\prettyref{app:Memory-curve}). The key idea
is to express the correlation functions $\langle x_{i}z\rangle$ as
a sum of response functions $\langle x_{i}\tx_{j}\rangle$; this is
possible due to the Gaussian statistics of $z$. The calculation is
similar to the derivation of the Schrödinger equation (\prettyref{app:pair_of_systems})
with the difference, however, that the two replicas receive independent
realizations of the inputs. The memory curve follows from a differential
equation for the correlation between the two systems and is measured
in units of the readout ratio $K/N$ \prettyref{eq:memory_final}:
\begin{eqnarray}
m(\tau) & = & \frac{2\sigma^{2}}{c_{0}}\,e^{-2\tau}\,I_{0}\left[2g\langle\phi^{\prime}(x)\rangle\,\tau\right]\,\Theta(\tau)\,d\tau\label{eq:memory_curve}
\end{eqnarray}
with the modified Bessel function of the first kind $I_{0}$. The
memory curve has two contributions \prettyref{eq:memory_final}: memory
due to the collective network dynamics and local memory due to the
leaky integration of the single units. The latter effect is trivial
and is reflected in the initial steep falloff $\propto e^{-2\tau}$
of the memory curves with time lag $\tau$, independent of the coupling
strength (\prettyref{fig:Memory-capacity}a). Its decay time is half
the time constant of the neurons, which is set to unity here \prettyref{eq:diffeq_motion}.
With increasing coupling strength the variance $c_{0}$ increases,
so that the memory curve \prettyref{eq:memory_curve} at zero time
lag $\tau=0$ reduces. For time lags that are large compared to the
single unit time constant, the network contribution to the memory
dominates. A non-vanishing memory capacity for longer time lags is
therefore only the result of the reverberation of the input through
the network interaction. The analytical results are in excellent agreement
with direct simulations.

We isolate the interesting network memory by subtracting the single
unit contribution (first term in \prettyref{eq:memory_final})
\begin{eqnarray}
m_{\text{net}}(\tau) & = & m(\tau)-\frac{2\sigma^{2}}{c_{0}}\,e^{-2\tau}\,\Theta(\tau)\,d\tau.\label{eq:memory_curve_net}
\end{eqnarray}
This quantity is particularly important in situations where the readout
does not have access to the neurons receiving the signal. The network
memory curve consistently vanishes for the uncoupled case. We compare
the performance of two different couplings strengths: $g_{\mathrm{\mathrm{\mathrm{nec}}}}$,
following from \prettyref{eq:local_instability} with equal sign and
corresponding to the onset of the local instability, and $g_{c}$,
marking the onset of chaos (cf. \prettyref{fig:Transition-to-chaos}b).
For short time lags, $\tau<5$, the network memory curve is larger
for $g_{\mathrm{\mathrm{\mathrm{nec}}}}$, while for longer time lags
it is larger for $g_{c}$ due to a slower decay of the memory curve.
This behavior is confirmed by the memory curve as a function of $g$,
shown for different time lags (\prettyref{fig:Memory-capacity}c,
upper panel). For $\tau\ge4$, the memory capacity $m$ is entirely
given by $m_{\text{net}}$, which is in line with the fast decay of
the single-unit contribution. Moreover, while for small time lags
the memory is maximal around $g_{\mathrm{\mathrm{nec}}}$, for larger
time lags it peaks nearby $g_{\mathrm{c}},$ indicating that the intermediate,
expansive, non-chaotic regime supports storage of the input.

The memory capacity is defined as the integral over the memory curve
\begin{eqnarray}
M & = & \int_{0}^{\infty}\,m(\tau)=\frac{\sigma^{2}}{c_{0}}\sqrt{\frac{1}{1-g^{2}\langle\phi^{\prime}(x)\rangle^{2}}},\label{eq:memory_capacity}
\end{eqnarray}
which follows directly from the Laplace transform of the Bessel function.

Typically the memory capacity is bounded by the number of neurons
$N$ \citep{Dambre12_514}. The signal $z$ in our situation, however,
can be seen as one out of $N$ independent inputs and $m(\tau)$ its
corresponding memory curve. The expressions are therefore independent
of $N$ \citep{Hermans10_1} and the memory capacity satisfies $M\le1$.
The network memory capacity is defined as $M_{\mathrm{net}}=\int\,m_{\text{net}}$.
While the memory capacity decreases in the chaotic regime, the network
memory peaks within the expansive, non-chaotic regime (\prettyref{fig:Memory-capacity}b,
lower panel).

\begin{figure}
\includegraphics{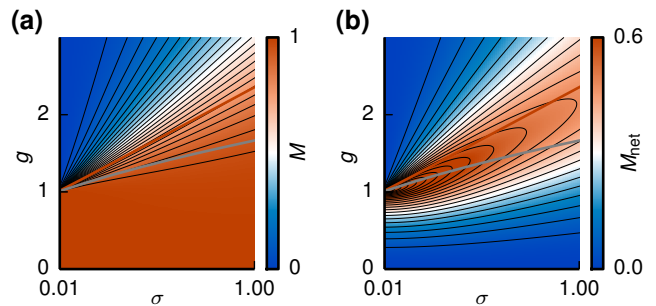}

\caption{\label{fig:mem_phasespace}\textbf{Memory capacity in different phases
of the network dynamics. (a) }Total memory capacity \prettyref{eq:memory_capacity}
encoded in color. Phase boundary \prettyref{eq:transition_crit} between
regular and chaotic dynamics (red) and necessary condition \prettyref{eq:local_instability}
of local instability (gray) as in \prettyref{fig:Transition-to-chaos}.
Contour lines of memory shown in black. \textbf{(b) }Same as (a) for
network contribution to memory capacity.}
\end{figure}

So far we have considered the memory capacity at a fixed amplitude
$\sigma$ of the input. In the following we investigate the memory
capacity over the whole phase diagram. The total memory capacity shows
a steep falloff directly above the onset of chaos (\prettyref{fig:mem_phasespace}a).
This is expected because the information about the input is lost in
the chaotic network dynamics. The contour lines of the memory capacity
are nearly parallel to the transition criterion, the curve with vanishing
Lyapunov exponent. This observation closely links the transition to
chaos to the memory capacity: The direction in the phase diagram in
which the system most quickly enters the chaotic regime is accompanied
by the steepest decline of memory. The contour lines of the total
network memory capacity show a ridge running through the expansive,
non-chaotic regime (\prettyref{fig:mem_phasespace}b); it confirms
the results found above: The memory is optimal in the dynamical regime
of local instability and asymptotic stability. Moreover, the network
capacity has a substantial contribution to the total memory capacity
of about $50\%$.

The optimal network memory in the hitherto unreported regime between
$g_{\mathrm{\mathrm{\mathrm{nec}}}}$ and $g_{\mathrm{c}}$ can be
understood in an intuitive manner. Two conditions must be met for
good memory. First, individual units must be susceptible to the signal;
the susceptibility equals the noise-averaged slope $\langle\phi^{\prime}\rangle$
of the gain function. Second, the signal must propagate effectively
through the network, requiring a sufficiently strong coupling $g$.
These two requirements are reflected in the monotonic increase of
the memory curve \prettyref{eq:memory_curve} with the effective slope
$g\,\langle\phi^{\prime}\rangle$, independent of the time lag $\tau$.
An increase in the coupling strength, however, elevates the intrinsically
generated fluctuations as well. These have a twofold effect on the
memory capacity. First they decrease $\langle\phi^{\prime}\rangle$,
so that $g\,\langle\phi^{\prime}\rangle$ assumes a maximum. Second,
the intrinsic fluctuations propagate through the network as well;
they hence reduce the signal to noise ratio of the readout, as they
enter the denominator in $M_{\mathrm{net}}$ through $c_{0}$. The
interplay of these two mechanisms leads to optimal memory located
in the expansive, non-chaotic regime, explaining why the combination
of time-local expansive dynamics and stable long-term behavior maximizes
the memory capacity of the network.

\section{Discussion}

We here present a completely solvable network model that allows us
to investigate the effect of time-varying input on the transition
to chaos and information processing capabilities. Adding time-varying
stochastic forcing to the seminal model by \citet{Sompolinsky88_259}
yields a stochastic continuous-time dynamical system. Contrary to
the original model \citep{Sompolinsky88_259}, we here reformulate
the stochastic differential equations as a field theory \citep{Martin73,dedominicis1976_247,janssen1976_377,DeDomincis78_353}.
This formal step allows us to develop the dynamic mean-field theory
by standard tools: a saddle point approximation of the auxiliary field
generating functional \citep{Sompolinsky82_6860,Crisanti87_4922,Moshe03}.
As in the original model, this procedure reduces the interacting system
to the dynamics of a single unit. The self-consistent solution of
the effective equation yields a standard physics problem: the autocorrelation
function of a typical unit is given by the motion of a classical particle
in a potential. We find that the amplitude of the input corresponds
to the initial kinetic energy of the particle. 

The field theoretical formulation then allows us to perform a replica
calculation to determine the maximum Lyapunov exponent; the problem
formally reduces to finding the ground-state energy of a single-particle
quantum mechanical problem. The transition to chaos appears at the
point where the ground state energy changes sign, which allows us
to obtain a closed form condition relating the coupling strength and
the input amplitude at the transition. We find a simple hallmark of
the transition in the single unit activity: at the transition point
the variance of the recurrent input to a single unit equals the variance
of its own activity. Correspondingly, the autocorrelation function
at zero time lag changes its curvature from convex to concave at the
transition point. These features can readily be measured in most physical
systems. The assessment of chaos by these passive observations in
particular does not require a perturbation of the system.

The transition criterion allows us to map out the phase diagram spanned
by the coupling strength and the input amplitude. It shows that external
drive shifts the transition to chaos to significantly larger coupling
strengths than predicted by time-local linear stability analysis.
The transition in the stochastic system is thus qualitatively different
from the transition in the autonomous system, where loss of local
stability and transition to chaos are equivalent \citep{Sompolinsky88_259,Sommers88}.
The discrepancy of these two measures in the driven system is explained
by a dynamic effect: the decrease of the maximum Lyapunov exponent
is related to the sharpening of the autocorrelation function by the
fluctuating drive. The displacement between local instability and
transition to chaos leads to an intermediate regime which is absent
in time-discrete networks \citep{molgedey92_3717} and only exists
in their more realistic time-continuous counterparts studied here.
This hitherto unreported dynamical regime combines locally expansive
dynamics with asymptotic stability. 

The seminal works \citep{Sompolinsky88_259,Sommers88} have established
a tight link between the fields of random matrix theory and autonomous
neural networks with random topology: Deterministic chaos emerges
if the spectral radius of the coupling matrix exceeds unity. In contrast,
we find in stochastically driven networks that the spectral radius
only yields a necessary condition for a positive Lyapunov exponent;
it determines the minimum of the quantum mechanical potential whose
ground state energy relates to the Lyapunov exponent. The presented
closed-form relation between input strength, the statistics of the
random matrix, and the onset of chaos \prettyref{eq:transition_crit}
generalizes the well known link to non-autonomous stochastic dynamics. 

It is controversially discussed whether the instability of deterministic
rate dynamics explains a transition to chaos in networks of spiking
neurons \citep{Ostojic14,engelken15_017798,ostojic15_020354}. It
was argued that such a transition is absent in spiking models because
the correlation time does not peak at the point where the corresponding
deterministic rate dynamics becomes unstable \citep{engelken15_017798,ostojic15_020354,Mastroguiseppe16_arxiv}.
For the analysis of oscillations \citep{Brunel99} and correlations
\citep{Trousdale12_e1002408,Helias13_023002} in these networks, the
irregular spiking activity of the neurons can be approximated by effective
stochastic rate equations, whereby the realization of the spikes is
represented by an explicit source of noise. In this setting, the input
$\xi$ in \eqref{eq:diffeq_motion} can be interpreted as such spiking
noise, explicitly investigated in \citep{Kadmon15_041030}. For weak
noise one may neglect its impact on the location of the transition.
In this limit, noise suppresses the divergence of the correlation
time \citep{Kadmon15_041030}. Our work is not bound to small noise
amplitudes and suggests that a diverging time scale in spiking networks
does not occur at the instability for two reasons. First, we have
shown that in a stochastic system the transition to chaos is not predicted
by local instability. If a diverging time-scale at the transition
existed it would not occur at the instability, but as a larger coupling
strength. But the presented analysis shows that the decay time of
the autocorrelation function does not even peak at the transition
to chaos, but rather in the chaotic regime. While these results strictly
only hold for the rate dynamics considered here, they still strongly
suggest the absence of a diverging time scale in networks of spiking
neurons. Indeed, the absence of a diverging time-scale has been observed
in simulation of spiking neurons \citep{engelken15_017798} as well
as in an iterative approach solving for the self-consistent autocorrelation
function \citep{dummer14,Wieland2015_040901}. 

To assess whether this richer dynamics found in the driven network
has functional consequences, we investigate sequential memory \citep{Jaeger01_memory}.
The obtained closed-form expression for the network memory capacity
exhibits a peak within the expansive, non-chaotic regime. We identify
two mechanisms whose partly antagonistic interplay causes optimal
memory: Local amplification of the stimulus and intrinsically-generated
noise. Local instability of the network ensures sensitivity to the
external input, so that on short time scales the incoming signal is
amplified and can therefore more reliably be read out. But larger
coupling also increases network-intrinsic fluctuations, which, in
turn, reduce the susceptibility as well as the signal to noise ratio
of the readout. Therefore it is plausible that the optimal memory
appears at a point where local amplification of the external input
is large enough, but intrinsic chaoticity is still limited.

Sequential memory has been studied in a time-discrete neural network
model \citep{Toyoizumi11_051908}, which receives a single weak external
input. In contrast, we here investigate the memory of a single signal
in the presence of multiple simultaneous inputs with arbitrary amplitude
$\sigma$. Without additional observation noise, \citet{Toyoizumi11_051908}
find that sequential memory does not posses a maximum; it is constant
and optimal for sub-critical coupling values $0<g<1$ and falls off
in the chaotic regime $g>1$ due to intrinsically-generated fluctuations.
Perfect reconstruction in the non-chaotic regime is possible, because
the single-step delayed activity is a direct linear function of the
input. In our setting, the single neuron memory has a similar effect
(\prettyref{fig:mem_phasespace} a). In the discrete system, optimal
memory close to the transition only arises in presence of observation
noise \citep{Toyoizumi11_051908}. Memory falloff in the chaotic phase
is much more shallow than in the direction of regular dynamics, so
that a fine-tuning is not needed if the network dynamic is slightly
chaotic.

For continuous dynamics the situation is qualitatively different.
The network component of the memory or, equivalently, the memory for
longer delay times $\tau$, shows non-monotonic behavior even without
observation noise. For small signal amplitude $\sigma$, memory is
optimal right below the transition to chaos and steeply falls off
above (\prettyref{fig:mem_phasespace} b). For large $\sigma$, the
falloff is weaker in the chaotic regime, qualitatively more similar
to \citet{Toyoizumi11_051908}.

A negative maximum Lyapunov exponent for nonautonomous dynamics indicates
the echo state property, the reliability of the network response to
input \citep{Wainrib16_39}. We could indeed show for the analytically
tractable model here that memory capacity quickly declines in the
chaotic regime due to intrinsically generated chaotic fluctuations.
Echo state networks show long temporal memory near the edge of chaos
\citep{Bertschinger04_1413,Legenstein07_127,Busing10_1272,Boedecker12_205}.
Typically these networks are time-discrete and thus the onset of chaos
is directly linked to the spectral radius of the Jacobian. This relation
is used in the design of these systems, exploiting that a spectral
radius close to local instability ensures long memory times. We here
show for the driven and time-continuous system, that the edge of chaos
and local instability are two different concepts and that memory capacity
is a third, distinct measure: Memory is optimal at neither of the
two other criteria, but rather in between. In particular, our analytical
results for the memory capacity can be used to determine the optimal
coupling strength for a given input amplitude.

Recently, an algorithm was proposed to train a random network as
given by \prettyref{eq:diffeq_motion} to produce a wide range of
activity patterns \citep{Sussillo09_544}. Such learning shows best
performance if initially the random network without input is in the
chaotic state. It has been argued that such networks have a large
dynamic range and are able to produce a wide variety of outputs. In
the training phase the input to the network needs to suppress chaos
so that learning converges. The procedure therefore requires the choice
of an initial coupling that is large enough to ensure chaos, but not
too large so that the input can suppress chaos.  Our quantitative
criterion for the transition easily enables a proper choice of parameters
and facilitates the design, control, and understanding of functional
networks.

In this work we have considered memory of the input signal. An important
task of the brain, however, is not only to maintain the input but
also to perform non-linear transformations on it. We expect the locally
unstable but globally stable dynamics to be beneficial for such a
task: The expansive behavior can project the input into a high-dimensional
space, which is crucial for non-linear computations or discrimination
tasks \citep{Legenstein07_127}. Thus, this dynamical regime not only
provides memory but might serve as a basis for more complex computations.

To show the generic effect of input we added a time-varying input
to the seminal model by \citet{Sompolinsky88_259}. Even though the
original model makes some simplifying assumptions, such as the all-to-all
Gaussian connectivity and a sigmoidal symmetric gain function, the
transition to chaos is qualitatively the same as in networks with
more biologically realistic parameters \citep{Mastroguiseppe16_arxiv,Kadmon15_041030,Harish15_e1004266}.
To focus on the new physics arising in non-autonomous systems, we
have here chosen to present the simplest but non-trivial, and yet
application-relevant extension. The reformulation of the derivation
of the dynamic mean-field theory by help of established methods from
field theory \citep{Sompolinsky82_6860,Crisanti87_4922,Moshe03,Chow15,Hertz16_033001}
here allowed us to find the explicit form of network memory by a replica
calculation. In general, the presented formulation opens the study
of recurrent random networks to the rich and powerful set of field
theoretical methods developed in other branches of physics. This language
allows a straight-forward extension of our results in various directions.
Among them more biologically realistic settings, such as sparse connectivity
respecting Dale's law \citep{Eccles54_524}, threshold like-activation
functions, non-negative activity variables or multiple populations.
The latter extension would allow the study of the interesting case
in which the population receiving the signal is separated from the
readout population. Such a situation would most likely emerge in the
cortex where the input population of local microcircuits typically
differs from the output population. 

More generally, the stability of complex dynamical systems plays an
important role in various other field of physics, biology, and technology.
Examples include oscillator networks \citep{Rodrigues16_1}, disordered
soft-spin models \citep{Sompolinsky81}, power grids \citep{Nishikawa15_015012},
food webs \citep{Allesina15_63}, and gene-regularity networks \citep{Pomerance09_8209}.
Presenting exact results for a prototypical and solvable model this
work contributes to the understanding of chaos and signal propagation
in such high dimensional systems.

\section{Acknowledgements}

This work was partially supported by Helmholtz young investigator's
group VH-NG-1028, Helmholtz portfolio theme SMHB, Jülich Aachen Research
Alliance (JARA). This project received funding from the European Union's
Horizon 2020 research and innovation programme under grant agreement
No. 720270. J.S. and S.G. contributed equally to this work. 

\bibliographystyle{apsrev_brain}

\begin{thebibliography}{69}
\expandafter\ifx\csname natexlab\endcsname\relax\def\natexlab#1{#1}\fi
\expandafter\ifx\csname bibnamefont\endcsname\relax
  \def\bibnamefont#1{#1}\fi
\expandafter\ifx\csname bibfnamefont\endcsname\relax
  \def\bibfnamefont#1{#1}\fi
\expandafter\ifx\csname citenamefont\endcsname\relax
  \def\citenamefont#1{#1}\fi
\expandafter\ifx\csname url\endcsname\relax
  \def\url#1{\texttt{#1}}\fi
\expandafter\ifx\csname urlprefix\endcsname\relax\def\urlprefix{URL }\fi
\providecommand{\bibinfo}[2]{#2}
\providecommand{\eprint}[2][]{\url{#2}}

\bibitem[{\citenamefont{Sompolinsky et~al.}(1988)\citenamefont{Sompolinsky,
  Crisanti, and Sommers}}]{Sompolinsky88_259}
\bibinfo{author}{\bibfnamefont{H.}~\bibnamefont{Sompolinsky}},
  \bibinfo{author}{\bibfnamefont{A.}~\bibnamefont{Crisanti}}, \bibnamefont{and}
  \bibinfo{author}{\bibfnamefont{H.~J.} \bibnamefont{Sommers}},
  \bibinfo{journal}{Phys. Rev. Lett.} \textbf{\bibinfo{volume}{61}},
  \bibinfo{pages}{259} (\bibinfo{year}{1988}).

\bibitem[{\citenamefont{van Vreeswijk and Sompolinsky}(1996)}]{Vreeswijk96}
\bibinfo{author}{\bibfnamefont{C.}~\bibnamefont{van Vreeswijk}}
  \bibnamefont{and}
  \bibinfo{author}{\bibfnamefont{H.}~\bibnamefont{Sompolinsky}},
  \bibinfo{journal}{Science} \textbf{\bibinfo{volume}{274}},
  \bibinfo{pages}{1724} (\bibinfo{year}{1996}).

\bibitem[{\citenamefont{Monteforte and Wolf}(2010)}]{Monteforte10_268104}
\bibinfo{author}{\bibfnamefont{M.}~\bibnamefont{Monteforte}} \bibnamefont{and}
  \bibinfo{author}{\bibfnamefont{F.}~\bibnamefont{Wolf}},
  \bibinfo{journal}{Phys. Rev. Lett.} \textbf{\bibinfo{volume}{105}},
  \bibinfo{pages}{268104} (\bibinfo{year}{2010}).

\bibitem[{\citenamefont{Lajoie et~al.}(2013)\citenamefont{Lajoie, Lin, and
  Shea-Brown}}]{Lajoie13_052901}
\bibinfo{author}{\bibfnamefont{G.}~\bibnamefont{Lajoie}},
  \bibinfo{author}{\bibfnamefont{K.~K.} \bibnamefont{Lin}}, \bibnamefont{and}
  \bibinfo{author}{\bibfnamefont{E.}~\bibnamefont{Shea-Brown}},
  \bibinfo{journal}{Phys. Rev. E} \textbf{\bibinfo{volume}{87}},
  \bibinfo{pages}{052901} (\bibinfo{year}{2013}).

\bibitem[{\citenamefont{Maass et~al.}(2002)\citenamefont{Maass,
  Natschl\"{a}ger, and Markram}}]{Maass02_2531}
\bibinfo{author}{\bibfnamefont{W.}~\bibnamefont{Maass}},
  \bibinfo{author}{\bibfnamefont{T.}~\bibnamefont{Natschl\"{a}ger}},
  \bibnamefont{and} \bibinfo{author}{\bibfnamefont{H.}~\bibnamefont{Markram}},
  \bibinfo{journal}{Neural Comput.} \textbf{\bibinfo{volume}{14}},
  \bibinfo{pages}{2531} (\bibinfo{year}{2002}).

\bibitem[{\citenamefont{Jaeger and Haas}(2004)}]{Jaeger04_87}
\bibinfo{author}{\bibfnamefont{H.}~\bibnamefont{Jaeger}} \bibnamefont{and}
  \bibinfo{author}{\bibfnamefont{H.}~\bibnamefont{Haas}},
  \bibinfo{journal}{Science} \textbf{\bibinfo{volume}{304}},
  \bibinfo{pages}{78} (\bibinfo{year}{2004}).

\bibitem[{\citenamefont{Legenstein and
  Maass}(2007{\natexlab{a}})}]{Legenstein07_323}
\bibinfo{author}{\bibfnamefont{R.}~\bibnamefont{Legenstein}} \bibnamefont{and}
  \bibinfo{author}{\bibfnamefont{W.}~\bibnamefont{Maass}},
  \bibinfo{journal}{Neural Networks} \textbf{\bibinfo{volume}{20}},
  \bibinfo{pages}{323} (\bibinfo{year}{2007}{\natexlab{a}}).

\bibitem[{\citenamefont{Sussillo and Abbott}(2009)}]{Sussillo09_544}
\bibinfo{author}{\bibfnamefont{D.}~\bibnamefont{Sussillo}} \bibnamefont{and}
  \bibinfo{author}{\bibfnamefont{L.~F.} \bibnamefont{Abbott}},
  \bibinfo{journal}{Neuron} \textbf{\bibinfo{volume}{63}}, \bibinfo{pages}{544}
  (\bibinfo{year}{2009}).

\bibitem[{\citenamefont{Toyoizumi and Abbott}(2011)}]{Toyoizumi11_051908}
\bibinfo{author}{\bibfnamefont{T.}~\bibnamefont{Toyoizumi}} \bibnamefont{and}
  \bibinfo{author}{\bibfnamefont{L.~F.} \bibnamefont{Abbott}},
  \bibinfo{journal}{Phys. Rev. E} \textbf{\bibinfo{volume}{84}},
  \bibinfo{pages}{051908} (\bibinfo{year}{2011}).

\bibitem[{\citenamefont{Barak et~al.}(2013)\citenamefont{Barak, Sussillo, Romo,
  Tsodyks, and Abbott}}]{Barak13_214}
\bibinfo{author}{\bibfnamefont{O.}~\bibnamefont{Barak}},
  \bibinfo{author}{\bibfnamefont{D.}~\bibnamefont{Sussillo}},
  \bibinfo{author}{\bibfnamefont{R.}~\bibnamefont{Romo}},
  \bibinfo{author}{\bibfnamefont{M.}~\bibnamefont{Tsodyks}}, \bibnamefont{and}
  \bibinfo{author}{\bibfnamefont{L.}~\bibnamefont{Abbott}},
  \textbf{\bibinfo{volume}{103}}, \bibinfo{pages}{214} (\bibinfo{year}{2013}).

\bibitem[{\citenamefont{Rajan et~al.}(2016)\citenamefont{Rajan, Harvey, and
  Tank}}]{Rajan16_128}
\bibinfo{author}{\bibfnamefont{K.}~\bibnamefont{Rajan}},
  \bibinfo{author}{\bibfnamefont{C.~D.} \bibnamefont{Harvey}},
  \bibnamefont{and} \bibinfo{author}{\bibfnamefont{D.~W.} \bibnamefont{Tank}},
  \bibinfo{journal}{Neuron} \textbf{\bibinfo{volume}{90}}, \bibinfo{pages}{128}
  (\bibinfo{year}{2016}).

\bibitem[{\citenamefont{Li et~al.}(2016)\citenamefont{Li, Daie, Svoboda, and
  Druckmann}}]{Li16_459}
\bibinfo{author}{\bibfnamefont{N.}~\bibnamefont{Li}},
  \bibinfo{author}{\bibfnamefont{K.}~\bibnamefont{Daie}},
  \bibinfo{author}{\bibfnamefont{K.}~\bibnamefont{Svoboda}}, \bibnamefont{and}
  \bibinfo{author}{\bibfnamefont{S.}~\bibnamefont{Druckmann}},
  \bibinfo{journal}{Nature} \textbf{\bibinfo{volume}{532}},
  \bibinfo{pages}{459} (\bibinfo{year}{2016}).

\bibitem[{\citenamefont{Murray et~al.}(2016)\citenamefont{Murray, Bernacchia,
  Roy, Constantinidis, Romo, and Wang}}]{Murray16_201619449}
\bibinfo{author}{\bibfnamefont{J.~D.} \bibnamefont{Murray}},
  \bibinfo{author}{\bibfnamefont{A.}~\bibnamefont{Bernacchia}},
  \bibinfo{author}{\bibfnamefont{N.~A.} \bibnamefont{Roy}},
  \bibinfo{author}{\bibfnamefont{C.}~\bibnamefont{Constantinidis}},
  \bibinfo{author}{\bibfnamefont{R.}~\bibnamefont{Romo}}, \bibnamefont{and}
  \bibinfo{author}{\bibfnamefont{X.-J.} \bibnamefont{Wang}},
  \bibinfo{journal}{Proc. Nat. Acad. Sci. USA} p. \bibinfo{pages}{201619449}
  (\bibinfo{year}{2016}).

\bibitem[{\citenamefont{Laje and Buonomano}(2013)}]{Laje13_925}
\bibinfo{author}{\bibfnamefont{R.}~\bibnamefont{Laje}} \bibnamefont{and}
  \bibinfo{author}{\bibfnamefont{D.~V.} \bibnamefont{Buonomano}},
  \bibinfo{journal}{Nat. Neurosci.} \textbf{\bibinfo{volume}{16}},
  \bibinfo{pages}{925} (\bibinfo{year}{2013}).

\bibitem[{\citenamefont{Mante et~al.}(2013)\citenamefont{Mante, Sussillo,
  Shenoy, and Newsome}}]{Mante13_78}
\bibinfo{author}{\bibfnamefont{V.}~\bibnamefont{Mante}},
  \bibinfo{author}{\bibfnamefont{D.}~\bibnamefont{Sussillo}},
  \bibinfo{author}{\bibfnamefont{K.~V.} \bibnamefont{Shenoy}},
  \bibnamefont{and} \bibinfo{author}{\bibfnamefont{W.~T.}
  \bibnamefont{Newsome}}, \bibinfo{journal}{Nature}
  \textbf{\bibinfo{volume}{503}}, \bibinfo{pages}{78} (\bibinfo{year}{2013}).

\bibitem[{\citenamefont{Sompolinsky and Zippelius}(1981)}]{Sompolinsky81}
\bibinfo{author}{\bibfnamefont{H.}~\bibnamefont{Sompolinsky}} \bibnamefont{and}
  \bibinfo{author}{\bibfnamefont{A.}~\bibnamefont{Zippelius}},
  \bibinfo{journal}{Phys. Rev. Lett.} \textbf{\bibinfo{volume}{47}},
  \bibinfo{pages}{359} (\bibinfo{year}{1981}).

\bibitem[{\citenamefont{Sompolinsky and Zippelius}(1982)}]{Sompolinsky82_6860}
\bibinfo{author}{\bibfnamefont{H.}~\bibnamefont{Sompolinsky}} \bibnamefont{and}
  \bibinfo{author}{\bibfnamefont{A.}~\bibnamefont{Zippelius}},
  \bibinfo{journal}{Phys. Rev. B} \textbf{\bibinfo{volume}{25}},
  \bibinfo{pages}{6860} (\bibinfo{year}{1982}).

\bibitem[{\citenamefont{Rajan et~al.}(2010)\citenamefont{Rajan, Abbott, and
  Sompolinsky}}]{rajan10_011903}
\bibinfo{author}{\bibfnamefont{K.}~\bibnamefont{Rajan}},
  \bibinfo{author}{\bibfnamefont{L.}~\bibnamefont{Abbott}}, \bibnamefont{and}
  \bibinfo{author}{\bibfnamefont{H.}~\bibnamefont{Sompolinsky}},
  \bibinfo{journal}{Phys. Rev. E} \textbf{\bibinfo{volume}{82}},
  \bibinfo{pages}{011903} (\bibinfo{year}{2010}).

\bibitem[{\citenamefont{Aljadeff et~al.}(2015)\citenamefont{Aljadeff, Stern,
  and Sharpee}}]{Aljadeff15_088101}
\bibinfo{author}{\bibfnamefont{J.}~\bibnamefont{Aljadeff}},
  \bibinfo{author}{\bibfnamefont{M.}~\bibnamefont{Stern}}, \bibnamefont{and}
  \bibinfo{author}{\bibfnamefont{T.}~\bibnamefont{Sharpee}},
  \bibinfo{journal}{Phys. Rev. Lett.} \textbf{\bibinfo{volume}{114}},
  \bibinfo{pages}{088101} (\bibinfo{year}{2015}).

\bibitem[{\citenamefont{Harish and Hansel}(2015)}]{Harish15_e1004266}
\bibinfo{author}{\bibfnamefont{O.}~\bibnamefont{Harish}} \bibnamefont{and}
  \bibinfo{author}{\bibfnamefont{D.}~\bibnamefont{Hansel}},
  \bibinfo{journal}{PLoS Comput Biol} \textbf{\bibinfo{volume}{11}},
  \bibinfo{pages}{e1004266} (\bibinfo{year}{2015}).

\bibitem[{\citenamefont{Sommers et~al.}(1988)\citenamefont{Sommers, Crisanti,
  Sompolinsky, and Stein}}]{Sommers88}
\bibinfo{author}{\bibfnamefont{H.}~\bibnamefont{Sommers}},
  \bibinfo{author}{\bibfnamefont{A.}~\bibnamefont{Crisanti}},
  \bibinfo{author}{\bibfnamefont{H.}~\bibnamefont{Sompolinsky}},
  \bibnamefont{and} \bibinfo{author}{\bibfnamefont{Y.}~\bibnamefont{Stein}},
  \bibinfo{journal}{Phys. Rev. Lett.} \textbf{\bibinfo{volume}{60}},
  \bibinfo{pages}{1895} (\bibinfo{year}{1988}).

\bibitem[{\citenamefont{Rajan and Abbott}(2006)}]{Rajan06}
\bibinfo{author}{\bibfnamefont{K.}~\bibnamefont{Rajan}} \bibnamefont{and}
  \bibinfo{author}{\bibfnamefont{L.~F.} \bibnamefont{Abbott}},
  \bibinfo{journal}{Phys. Rev. Lett.} \textbf{\bibinfo{volume}{97}},
  \bibinfo{pages}{188104} (\bibinfo{year}{2006}).

\bibitem[{\citenamefont{Molgedey et~al.}(1992)\citenamefont{Molgedey,
  Schuchhardt, and Schuster}}]{molgedey92_3717}
\bibinfo{author}{\bibfnamefont{L.}~\bibnamefont{Molgedey}},
  \bibinfo{author}{\bibfnamefont{J.}~\bibnamefont{Schuchhardt}},
  \bibnamefont{and} \bibinfo{author}{\bibfnamefont{H.}~\bibnamefont{Schuster}},
  \bibinfo{journal}{Phys. Rev. Lett.} \textbf{\bibinfo{volume}{69}},
  \bibinfo{pages}{3717} (\bibinfo{year}{1992}).

\bibitem[{\citenamefont{Zhou and Kurths}(2002)}]{Zhou02_230602}
\bibinfo{author}{\bibfnamefont{C.}~\bibnamefont{Zhou}} \bibnamefont{and}
  \bibinfo{author}{\bibfnamefont{J.}~\bibnamefont{Kurths}},
  \bibinfo{journal}{Phys. Rev. Lett.} \textbf{\bibinfo{volume}{88}},
  \bibinfo{pages}{230602} (\bibinfo{year}{2002}).

\bibitem[{\citenamefont{Martin et~al.}(1973)\citenamefont{Martin, Siggia, and
  Rose}}]{Martin73}
\bibinfo{author}{\bibfnamefont{P.}~\bibnamefont{Martin}},
  \bibinfo{author}{\bibfnamefont{E.}~\bibnamefont{Siggia}}, \bibnamefont{and}
  \bibinfo{author}{\bibfnamefont{H.}~\bibnamefont{Rose}},
  \bibinfo{journal}{Phys. Rev. A} \textbf{\bibinfo{volume}{8}},
  \bibinfo{pages}{423} (\bibinfo{year}{1973}).

\bibitem[{\citenamefont{Janssen}(1976)}]{janssen1976_377}
\bibinfo{author}{\bibfnamefont{H.-K.} \bibnamefont{Janssen}},
  \bibinfo{journal}{Zeitschrift f{\"u}r Physik B Condensed Matter}
  \textbf{\bibinfo{volume}{23}}, \bibinfo{pages}{377} (\bibinfo{year}{1976}).

\bibitem[{\citenamefont{De~Dominicis}(1976)}]{dedominicis1976_247}
\bibinfo{author}{\bibfnamefont{C.}~\bibnamefont{De~Dominicis}},
  \bibinfo{journal}{J. Phys. Colloques} \textbf{\bibinfo{volume}{37}},
  \bibinfo{pages}{C1} (\bibinfo{year}{1976}).

\bibitem[{\citenamefont{De~Dominicis and Peliti}(1978)}]{DeDomincis78_353}
\bibinfo{author}{\bibfnamefont{C.}~\bibnamefont{De~Dominicis}}
  \bibnamefont{and} \bibinfo{author}{\bibfnamefont{L.}~\bibnamefont{Peliti}},
  \bibinfo{journal}{Phys. Rev. B} \textbf{\bibinfo{volume}{18}},
  \bibinfo{pages}{353} (\bibinfo{year}{1978}).

\bibitem[{\citenamefont{Schuecker et~al.}(2016)\citenamefont{Schuecker,
  Goedeke, Dahmen, and Helias}}]{Schuecker16b_arxiv}
\bibinfo{author}{\bibfnamefont{J.}~\bibnamefont{Schuecker}},
  \bibinfo{author}{\bibfnamefont{S.}~\bibnamefont{Goedeke}},
  \bibinfo{author}{\bibfnamefont{D.}~\bibnamefont{Dahmen}}, \bibnamefont{and}
  \bibinfo{author}{\bibfnamefont{M.}~\bibnamefont{Helias}},
  \bibinfo{journal}{arXiv}  (\bibinfo{year}{2016}), \bibinfo{note}{1605.06758
  [cond-mat.dis-nn]}.

\bibitem[{\citenamefont{Moshe and Zinn-Justin}(2003)}]{Moshe03}
\bibinfo{author}{\bibfnamefont{M.}~\bibnamefont{Moshe}} \bibnamefont{and}
  \bibinfo{author}{\bibfnamefont{J.}~\bibnamefont{Zinn-Justin}},
  \bibinfo{journal}{Physics Reports} \textbf{\bibinfo{volume}{385}},
  \bibinfo{pages}{69} (\bibinfo{year}{2003}), ISSN \bibinfo{issn}{0370-1573}.

\bibitem[{\citenamefont{Eckmann and Ruelle}(1985)}]{Eckmann85}
\bibinfo{author}{\bibfnamefont{J.-P.} \bibnamefont{Eckmann}} \bibnamefont{and}
  \bibinfo{author}{\bibfnamefont{D.}~\bibnamefont{Ruelle}},
  \bibinfo{journal}{Reviews of modern physics} \textbf{\bibinfo{volume}{57}},
  \bibinfo{pages}{617} (\bibinfo{year}{1985}).

\bibitem[{\citenamefont{Derrida and Pomeau}(1986)}]{Derrida86_45}
\bibinfo{author}{\bibfnamefont{B.}~\bibnamefont{Derrida}} \bibnamefont{and}
  \bibinfo{author}{\bibfnamefont{Y.}~\bibnamefont{Pomeau}},
  \bibinfo{journal}{EPL (Europhysics Letters)} \textbf{\bibinfo{volume}{1}},
  \bibinfo{pages}{45} (\bibinfo{year}{1986}).

\bibitem[{\citenamefont{Jaeger}(2001)}]{Jaeger01_memory}
\bibinfo{author}{\bibfnamefont{H.}~\bibnamefont{Jaeger}},
  \emph{\bibinfo{title}{Short term memory in echo state networks}},
  vol.~\bibinfo{volume}{5} (\bibinfo{publisher}{GMD-Forschungszentrum
  Informationstechnik}, \bibinfo{year}{2001}).

\bibitem[{\citenamefont{Fischer and Hertz}(1991)}]{Fischer91}
\bibinfo{author}{\bibfnamefont{K.}~\bibnamefont{Fischer}} \bibnamefont{and}
  \bibinfo{author}{\bibfnamefont{J.}~\bibnamefont{Hertz}},
  \emph{\bibinfo{title}{Spin glasses}} (\bibinfo{publisher}{Cambridge
  University Press}, \bibinfo{year}{1991}).

\bibitem[{\citenamefont{Amari}(1972)}]{Amari72_643}
\bibinfo{author}{\bibfnamefont{S.-I.} \bibnamefont{Amari}},
  \bibinfo{journal}{Systems, Man and Cybernetics, IEEE Transactions on} pp.
  \bibinfo{pages}{643--657} (\bibinfo{year}{1972}).

\bibitem[{\citenamefont{Crisanti and Sompolinsky}(1987)}]{Crisanti87_4922}
\bibinfo{author}{\bibfnamefont{A.}~\bibnamefont{Crisanti}} \bibnamefont{and}
  \bibinfo{author}{\bibfnamefont{H.}~\bibnamefont{Sompolinsky}},
  \bibinfo{journal}{Phys. Rev. A} \textbf{\bibinfo{volume}{36}},
  \bibinfo{pages}{4922} (\bibinfo{year}{1987}).

\bibitem[{\citenamefont{Zinn-Justin}(1996)}]{ZinnJustin96}
\bibinfo{author}{\bibfnamefont{J.}~\bibnamefont{Zinn-Justin}},
  \emph{\bibinfo{title}{Quantum field theory and critical phenomena}}
  (\bibinfo{publisher}{Clarendon Press, Oxford}, \bibinfo{year}{1996}).

\bibitem[{\citenamefont{Cabana and Touboul}(2013)}]{Cabana13_211}
\bibinfo{author}{\bibfnamefont{T.}~\bibnamefont{Cabana}} \bibnamefont{and}
  \bibinfo{author}{\bibfnamefont{J.}~\bibnamefont{Touboul}},
  \bibinfo{journal}{J. statist. Phys.} \textbf{\bibinfo{volume}{153}},
  \bibinfo{pages}{211} (\bibinfo{year}{2013}).

\bibitem[{\citenamefont{Gardiner}(1985)}]{Gardiner85}
\bibinfo{author}{\bibfnamefont{C.~W.} \bibnamefont{Gardiner}},
  \emph{\bibinfo{title}{Handbook of Stochastic Methods for Physics, Chemistry
  and the Natural Sciences}} (\bibinfo{publisher}{Springer-Verlag},
  \bibinfo{address}{Berlin}, \bibinfo{year}{1985}), \bibinfo{edition}{2nd} ed.,
  ISBN \bibinfo{isbn}{3-540-61634-9, 3-540-15607-0}.

\bibitem[{\citenamefont{Altland and B.}(2010)}]{Altland01}
\bibinfo{author}{\bibfnamefont{A.}~\bibnamefont{Altland}} \bibnamefont{and}
  \bibinfo{author}{\bibfnamefont{S.}~\bibnamefont{B.}},
  \emph{\bibinfo{title}{Concepts of Theoretical Solid State Physics}}
  (\bibinfo{publisher}{Cambridge university press}, \bibinfo{year}{2010}).

\bibitem[{\citenamefont{Kadmon and Sompolinsky}(2015)}]{Kadmon15_041030}
\bibinfo{author}{\bibfnamefont{J.}~\bibnamefont{Kadmon}} \bibnamefont{and}
  \bibinfo{author}{\bibfnamefont{H.}~\bibnamefont{Sompolinsky}},
  \bibinfo{journal}{Phys. Rev. X} \textbf{\bibinfo{volume}{5}},
  \bibinfo{pages}{041030} (\bibinfo{year}{2015}).

\bibitem[{\citenamefont{Price}(1958)}]{Price58_69}
\bibinfo{author}{\bibfnamefont{R.}~\bibnamefont{Price}}, \bibinfo{journal}{IRE
  Transactions on Information Theory} \textbf{\bibinfo{volume}{4}},
  \bibinfo{pages}{69} (\bibinfo{year}{1958}).

\bibitem[{\citenamefont{Papoulis}(1991)}]{Papoulis91}
\bibinfo{author}{\bibfnamefont{A.}~\bibnamefont{Papoulis}},
  \emph{\bibinfo{title}{Probability, Random Variables, and Stochastic
  Processes}} (\bibinfo{publisher}{McGraw-Hill}, \bibinfo{address}{Boston,
  Massachusetts}, \bibinfo{year}{1991}), \bibinfo{edition}{3rd} ed.

\bibitem[{Note1()}]{Note1}
Note1, \bibinfo{note}{the theory of random dynamical systems makes this more
  precise; a brief overview is given in \protect \citep {Lajoie13_052901}.}

\bibitem[{\citenamefont{Dambre et~al.}(2012)\citenamefont{Dambre, Verstraeten,
  Schrauwen, and Massar}}]{Dambre12_514}
\bibinfo{author}{\bibfnamefont{J.}~\bibnamefont{Dambre}},
  \bibinfo{author}{\bibfnamefont{D.}~\bibnamefont{Verstraeten}},
  \bibinfo{author}{\bibfnamefont{B.}~\bibnamefont{Schrauwen}},
  \bibnamefont{and} \bibinfo{author}{\bibfnamefont{S.}~\bibnamefont{Massar}},
  \bibinfo{journal}{Scientific reports} \textbf{\bibinfo{volume}{2}},
  \bibinfo{pages}{514} (\bibinfo{year}{2012}).

\bibitem[{\citenamefont{Hermans and Schrauwen}(2010)}]{Hermans10_1}
\bibinfo{author}{\bibfnamefont{M.}~\bibnamefont{Hermans}} \bibnamefont{and}
  \bibinfo{author}{\bibfnamefont{B.}~\bibnamefont{Schrauwen}}, in
  \emph{\bibinfo{booktitle}{Neural Networks (IJCNN), The 2010 International
  Joint Conference on}} (\bibinfo{organization}{IEEE}, \bibinfo{year}{2010}),
  pp. \bibinfo{pages}{1--7}.

\bibitem[{\citenamefont{Ostojic}(2014)}]{Ostojic14}
\bibinfo{author}{\bibfnamefont{S.}~\bibnamefont{Ostojic}},
  \bibinfo{journal}{Nat. Neurosci.} \textbf{\bibinfo{volume}{17}},
  \bibinfo{pages}{594} (\bibinfo{year}{2014}).

\bibitem[{\citenamefont{Engelken et~al.}(2015)\citenamefont{Engelken, Farkhooi,
  Hansel, van Vreeswijk, and Wolf}}]{engelken15_017798}
\bibinfo{author}{\bibfnamefont{R.}~\bibnamefont{Engelken}},
  \bibinfo{author}{\bibfnamefont{F.}~\bibnamefont{Farkhooi}},
  \bibinfo{author}{\bibfnamefont{D.}~\bibnamefont{Hansel}},
  \bibinfo{author}{\bibfnamefont{C.}~\bibnamefont{van Vreeswijk}},
  \bibnamefont{and} \bibinfo{author}{\bibfnamefont{F.}~\bibnamefont{Wolf}},
  \bibinfo{journal}{bioRxiv} p. \bibinfo{pages}{017798} (\bibinfo{year}{2015}).

\bibitem[{\citenamefont{Ostojic}(2015)}]{ostojic15_020354}
\bibinfo{author}{\bibfnamefont{S.}~\bibnamefont{Ostojic}},
  \bibinfo{journal}{bioRxiv} p. \bibinfo{pages}{020354} (\bibinfo{year}{2015}).

\bibitem[{\citenamefont{Mastroguiseppe and
  Ostojic}(2016)}]{Mastroguiseppe16_arxiv}
\bibinfo{author}{\bibfnamefont{F.}~\bibnamefont{Mastroguiseppe}}
  \bibnamefont{and} \bibinfo{author}{\bibfnamefont{S.}~\bibnamefont{Ostojic}},
  \bibinfo{journal}{arXiv} p. \bibinfo{pages}{1605.04221}
  (\bibinfo{year}{2016}).

\bibitem[{\citenamefont{Brunel and Hakim}(1999)}]{Brunel99}
\bibinfo{author}{\bibfnamefont{N.}~\bibnamefont{Brunel}} \bibnamefont{and}
  \bibinfo{author}{\bibfnamefont{V.}~\bibnamefont{Hakim}},
  \bibinfo{journal}{Neural Comput.} \textbf{\bibinfo{volume}{11}},
  \bibinfo{pages}{1621} (\bibinfo{year}{1999}).

\bibitem[{\citenamefont{Trousdale et~al.}(2012)\citenamefont{Trousdale, Hu,
  Shea-Brown, and Josic}}]{Trousdale12_e1002408}
\bibinfo{author}{\bibfnamefont{J.}~\bibnamefont{Trousdale}},
  \bibinfo{author}{\bibfnamefont{Y.}~\bibnamefont{Hu}},
  \bibinfo{author}{\bibfnamefont{E.}~\bibnamefont{Shea-Brown}},
  \bibnamefont{and} \bibinfo{author}{\bibfnamefont{K.}~\bibnamefont{Josic}},
  \bibinfo{journal}{PLOS Comput. Biol.} \textbf{\bibinfo{volume}{8}},
  \bibinfo{pages}{e1002408} (\bibinfo{year}{2012}).

\bibitem[{\citenamefont{Helias et~al.}(2013)\citenamefont{Helias, Tetzlaff, and
  Diesmann}}]{Helias13_023002}
\bibinfo{author}{\bibfnamefont{M.}~\bibnamefont{Helias}},
  \bibinfo{author}{\bibfnamefont{T.}~\bibnamefont{Tetzlaff}}, \bibnamefont{and}
  \bibinfo{author}{\bibfnamefont{M.}~\bibnamefont{Diesmann}},
  \bibinfo{journal}{New J. Phys.} \textbf{\bibinfo{volume}{15}},
  \bibinfo{pages}{023002} (\bibinfo{year}{2013}).

\bibitem[{\citenamefont{Dummer et~al.}(2014)\citenamefont{Dummer, Wieland, and
  Lindner}}]{dummer14}
\bibinfo{author}{\bibfnamefont{B.}~\bibnamefont{Dummer}},
  \bibinfo{author}{\bibfnamefont{S.}~\bibnamefont{Wieland}}, \bibnamefont{and}
  \bibinfo{author}{\bibfnamefont{B.}~\bibnamefont{Lindner}},
  \bibinfo{journal}{Front. Comput. Neurosci.} \textbf{\bibinfo{volume}{8}},
  \bibinfo{pages}{104} (\bibinfo{year}{2014}).

\bibitem[{\citenamefont{Wieland et~al.}(2015)\citenamefont{Wieland, Bernardi,
  Schwalger, and Lindner}}]{Wieland2015_040901}
\bibinfo{author}{\bibfnamefont{S.}~\bibnamefont{Wieland}},
  \bibinfo{author}{\bibfnamefont{D.}~\bibnamefont{Bernardi}},
  \bibinfo{author}{\bibfnamefont{T.}~\bibnamefont{Schwalger}},
  \bibnamefont{and} \bibinfo{author}{\bibfnamefont{B.}~\bibnamefont{Lindner}},
  \bibinfo{journal}{Phys. Rev. E} \textbf{\bibinfo{volume}{92}},
  \bibinfo{pages}{040901} (\bibinfo{year}{2015}).

\bibitem[{\citenamefont{Wainrib and Galtier}(2016)}]{Wainrib16_39}
\bibinfo{author}{\bibfnamefont{G.}~\bibnamefont{Wainrib}} \bibnamefont{and}
  \bibinfo{author}{\bibfnamefont{M.~N.} \bibnamefont{Galtier}},
  \bibinfo{journal}{Neural Networks} \textbf{\bibinfo{volume}{76}},
  \bibinfo{pages}{39 } (\bibinfo{year}{2016}), ISSN \bibinfo{issn}{0893-6080}.

\bibitem[{\citenamefont{Bertschinger and
  Natschl{\"a}ger}(2004)}]{Bertschinger04_1413}
\bibinfo{author}{\bibfnamefont{N.}~\bibnamefont{Bertschinger}}
  \bibnamefont{and}
  \bibinfo{author}{\bibfnamefont{T.}~\bibnamefont{Natschl{\"a}ger}},
  \bibinfo{journal}{Neural Comput.} \textbf{\bibinfo{volume}{16}},
  \bibinfo{pages}{1413} (\bibinfo{year}{2004}).

\bibitem[{\citenamefont{Legenstein and
  Maass}(2007{\natexlab{b}})}]{Legenstein07_127}
\bibinfo{author}{\bibfnamefont{R.}~\bibnamefont{Legenstein}} \bibnamefont{and}
  \bibinfo{author}{\bibfnamefont{W.}~\bibnamefont{Maass}},
  \emph{\bibinfo{title}{What makes a dynamical system computationally
  powerful?}} (\bibinfo{publisher}{MIT Press},
  \bibinfo{year}{2007}{\natexlab{b}}), pp. \bibinfo{pages}{127--154}.

\bibitem[{\citenamefont{B{\"u}sing et~al.}(2010)\citenamefont{B{\"u}sing,
  Schrauwen, and Legenstein}}]{Busing10_1272}
\bibinfo{author}{\bibfnamefont{L.}~\bibnamefont{B{\"u}sing}},
  \bibinfo{author}{\bibfnamefont{B.}~\bibnamefont{Schrauwen}},
  \bibnamefont{and}
  \bibinfo{author}{\bibfnamefont{R.}~\bibnamefont{Legenstein}},
  \bibinfo{journal}{Neural Comput.} \textbf{\bibinfo{volume}{22}},
  \bibinfo{pages}{1272} (\bibinfo{year}{2010}).

\bibitem[{\citenamefont{Boedecker et~al.}(2012)\citenamefont{Boedecker, Obst,
  Lizier, Mayer, and Asada}}]{Boedecker12_205}
\bibinfo{author}{\bibfnamefont{J.}~\bibnamefont{Boedecker}},
  \bibinfo{author}{\bibfnamefont{O.}~\bibnamefont{Obst}},
  \bibinfo{author}{\bibfnamefont{J.~T.} \bibnamefont{Lizier}},
  \bibinfo{author}{\bibfnamefont{N.~M.} \bibnamefont{Mayer}}, \bibnamefont{and}
  \bibinfo{author}{\bibfnamefont{M.}~\bibnamefont{Asada}},
  \bibinfo{journal}{Theory in Biosciences} \textbf{\bibinfo{volume}{131}},
  \bibinfo{pages}{205} (\bibinfo{year}{2012}).

\bibitem[{\citenamefont{Chow and Buice}(2015)}]{Chow15}
\bibinfo{author}{\bibfnamefont{C.}~\bibnamefont{Chow}} \bibnamefont{and}
  \bibinfo{author}{\bibfnamefont{M.}~\bibnamefont{Buice}},
  \bibinfo{journal}{The Journal of Mathematical Neuroscience}
  \textbf{\bibinfo{volume}{5}} (\bibinfo{year}{2015}).

\bibitem[{\citenamefont{Hertz et~al.}(2017)\citenamefont{Hertz, Roudi, and
  Sollich}}]{Hertz16_033001}
\bibinfo{author}{\bibfnamefont{J.~A.} \bibnamefont{Hertz}},
  \bibinfo{author}{\bibfnamefont{Y.}~\bibnamefont{Roudi}}, \bibnamefont{and}
  \bibinfo{author}{\bibfnamefont{P.}~\bibnamefont{Sollich}},
  \bibinfo{journal}{Journal of Physics A: Mathematical and Theoretical}
  \textbf{\bibinfo{volume}{50}}, \bibinfo{pages}{033001}
  (\bibinfo{year}{2017}).

\bibitem[{\citenamefont{Eccles et~al.}(1954)\citenamefont{Eccles, P, and
  Koketsu}}]{Eccles54_524}
\bibinfo{author}{\bibfnamefont{J.}~\bibnamefont{Eccles}},
  \bibinfo{author}{\bibfnamefont{F.}~\bibnamefont{P}}, \bibnamefont{and}
  \bibinfo{author}{\bibfnamefont{K.}~\bibnamefont{Koketsu}},
  \bibinfo{journal}{J Physiol (Lond)} \textbf{\bibinfo{volume}{126}},
  \bibinfo{pages}{524} (\bibinfo{year}{1954}).

\bibitem[{\citenamefont{Rodrigues et~al.}(2016)\citenamefont{Rodrigues, Peron,
  Ji, and Kurths}}]{Rodrigues16_1}
\bibinfo{author}{\bibfnamefont{F.~A.} \bibnamefont{Rodrigues}},
  \bibinfo{author}{\bibfnamefont{T.~K.~D.} \bibnamefont{Peron}},
  \bibinfo{author}{\bibfnamefont{P.}~\bibnamefont{Ji}}, \bibnamefont{and}
  \bibinfo{author}{\bibfnamefont{J.}~\bibnamefont{Kurths}},
  \bibinfo{journal}{Phys. Rep.} \textbf{\bibinfo{volume}{610}},
  \bibinfo{pages}{1} (\bibinfo{year}{2016}).

\bibitem[{\citenamefont{Nishikawa and Motter}(2015)}]{Nishikawa15_015012}
\bibinfo{author}{\bibfnamefont{T.}~\bibnamefont{Nishikawa}} \bibnamefont{and}
  \bibinfo{author}{\bibfnamefont{A.~E.} \bibnamefont{Motter}},
  \bibinfo{journal}{New J. Phys.} \textbf{\bibinfo{volume}{17}},
  \bibinfo{pages}{015012} (\bibinfo{year}{2015}).

\bibitem[{\citenamefont{Allesina and Tang}(2015)}]{Allesina15_63}
\bibinfo{author}{\bibfnamefont{S.}~\bibnamefont{Allesina}} \bibnamefont{and}
  \bibinfo{author}{\bibfnamefont{S.}~\bibnamefont{Tang}},
  \bibinfo{journal}{Population Ecology} \textbf{\bibinfo{volume}{57}},
  \bibinfo{pages}{63} (\bibinfo{year}{2015}).

\bibitem[{\citenamefont{Pomerance et~al.}(2009)\citenamefont{Pomerance, Ott,
  Girvan, and Losert}}]{Pomerance09_8209}
\bibinfo{author}{\bibfnamefont{A.}~\bibnamefont{Pomerance}},
  \bibinfo{author}{\bibfnamefont{E.}~\bibnamefont{Ott}},
  \bibinfo{author}{\bibfnamefont{M.}~\bibnamefont{Girvan}}, \bibnamefont{and}
  \bibinfo{author}{\bibfnamefont{W.}~\bibnamefont{Losert}},
  \bibinfo{journal}{Proc. Nat. Acad. Sci. USA} \textbf{\bibinfo{volume}{106}},
  \bibinfo{pages}{8209} (\bibinfo{year}{2009}).

\bibitem[{\citenamefont{Negele and Orland}(1998)}]{NegeleOrland98}
\bibinfo{author}{\bibfnamefont{J.~W.} \bibnamefont{Negele}} \bibnamefont{and}
  \bibinfo{author}{\bibfnamefont{H.}~\bibnamefont{Orland}},
  \emph{\bibinfo{title}{Quantum Many-Particle Systems}}
  (\bibinfo{publisher}{Perseus Books}, \bibinfo{year}{1998}).

\bibitem[{\citenamefont{Abramowitz and Stegun}(1974)}]{Abramowitz74}
\bibinfo{author}{\bibfnamefont{M.}~\bibnamefont{Abramowitz}} \bibnamefont{and}
  \bibinfo{author}{\bibfnamefont{I.~A.} \bibnamefont{Stegun}},
  \emph{\bibinfo{title}{Handbook of Mathematical Functions: with Formulas,
  Graphs, and Mathematical Tables}} (\bibinfo{publisher}{Dover Publications},
  \bibinfo{address}{New York}, \bibinfo{year}{1974}).

\end{thebibliography}

\global\long\def\erf{\mathrm{erf}}
\global\long\def\LN{\mathcal{L}_{0}}
\global\long\def\LO{\mathcal{L}_{1}}
\global\long\def\mV{\:\mathrm{mV}}
\global\long\def\ms{\:\mathrm{ms}}
\global\long\def\Hz{\:\mathrm{Hz}}
\global\long\def\D{\mathcal{D}}
\global\long\def\J{\mathbf{J}}
\global\long\def\one{\mathbf{1}}
\global\long\def\e{\mathbf{e}}
\global\long\def\Cpp{\mathcal{K}_{\phi^{\prime}\phi^{\prime}}^{(0)}}
\global\long\def\CCpp{C_{\phi^{\prime}\phi^{\prime}}^{(0)}}
\global\long\def\Cppj{\mathcal{K}_{\phi_{j}^{\prime}\phi_{j}^{\prime}}^{(0)}}
\global\long\def\tx{\tilde{x}}
\global\long\def\xo{x^{(0)}}
\global\long\def\xii{x^{(1)}}
\global\long\def\txi{\tilde{x}^{(1)}}
\global\long\def\bx{\mathbf{x}}
\global\long\def\tbx{\tilde{\mathbf{x}}}
\global\long\def\bl{\mathbf{l}}
\global\long\def\bh{\mathbf{h}}
\global\long\def\bJ{\mathbf{J}}
\global\long\def\bN{\mathcal{N}}
\global\long\def\bH{\mathbf{H}}
\global\long\def\bK{\mathbf{K}}
\global\long\def\bxo{\bx^{(0)}}
\global\long\def\tbxo{\tilde{\bx}^{(0)}}
\global\long\def\bxi{\bx^{(1)}}
\global\long\def\tbxi{\tilde{\bx}^{(1)}}
\global\long\def\tbxi{\tbx^{(1)}}
\global\long\def\tpsi{\tilde{\psi}}
\global\long\def\Cxi{C_{x^{(1)}x^{(1)}}}
\global\long\def\bxxi{\mathbf{\xi}}
\global\long\def\N{\mathcal{N}}
\global\long\def\bW{\mathbf{W}}
\global\long\def\bon{\mathbf{1}}
\global\long\def\unity{\mathds{1}}

\global\long\def\D{\mathcal{D}}
\global\long\def\J{\mathbf{J}}
\global\long\def\one{\mathbf{1}}
\global\long\def\e{\mathbf{e}}
\global\long\def\Cpp{\mathcal{K}_{\phi^{\prime}\phi^{\prime}}^{(0)}}
\global\long\def\CCpp{C_{\phi^{\prime}\phi^{\prime}}^{(0)}}
\global\long\def\Cppj{\mathcal{K}_{\phi_{j}^{\prime}\phi_{j}^{\prime}}^{(0)}}
\global\long\def\tx{\tilde{x}}
\global\long\def\xo{x^{(0)}}
\global\long\def\xii{x^{(1)}}
\global\long\def\txi{\tilde{x}^{(1)}}
\global\long\def\bx{\mathbf{x}}
\global\long\def\tbx{\tilde{\mathbf{x}}}
\global\long\def\bl{\mathbf{j}}
\global\long\def\tbj{\tilde{\mathbf{j}}}
\global\long\def\bk{\mathbf{k}}
\global\long\def\tbk{\tilde{\mathbf{k}}}
\global\long\def\bh{\mathbf{h}}
\global\long\def\bJ{\mathbf{J}}
\global\long\def\bN{\mathcal{N}}
\global\long\def\bH{\mathbf{H}}
\global\long\def\bK{\mathbf{K}}
\global\long\def\bxo{\bx^{(0)}}
\global\long\def\tbxo{\tilde{\bx}^{(0)}}
\global\long\def\bxi{\bx^{(1)}}
\global\long\def\tbxi{\tilde{\bx}^{(1)}}
\global\long\def\tbxi{\tbx^{(1)}}
\global\long\def\tpsi{\tilde{\psi}}
\global\long\def\Cxi{C_{x^{(1)}x^{(1)}}}
\global\long\def\bxxi{\mathbf{\xi}}
\global\long\def\N{\mathcal{N}}
\global\long\def\bW{\mathbf{W}}
\global\long\def\bon{\mathbf{1}}
\global\long\def\tj{\tilde{j}}
\global\long\def\tJ{\tilde{J}}
\global\long\def\Z{\mathcal{Z}}
\global\long\def\SOM{S_{\mathrm{OM}}}
\global\long\def\SMSR{S_{\mathrm{MSR}}}

\global\long\def\cX{\mathcal{X}}
\global\long\def\cK{\mathcal{K}}
\global\long\def\cF{\mathcal{F}}
\global\long\def\unity{\mathds{1}}
\global\long\def\bj{\mathbf{j}}
\global\long\def\bh{\mathbf{h}}

\global\long\def\D{\mathcal{D}}
\global\long\def\bx{\mathbf{x}}
\global\long\def\bl{\mathbf{l}}
\global\long\def\bh{\mathbf{h}}
\global\long\def\bJ{\mathbf{J}}
\global\long\def\N{\mathcal{N}}
\global\long\def\hh{\hat{h}}
\global\long\def\bhh{\mathbf{\hh}}
\global\long\def\T{\mathrm{T}}
\global\long\def\by{\mathrm{\mathbf{y}}}
\global\long\def\diag{\mathrm{diag}}
\global\long\def\Ftr#1#2{\mathcal{F}\left[#1\right]\left(#2\right)}
\global\long\def\iFtr#1#2{\mathfrak{\mathcal{F}^{-1}}\left[#1\right]\left(#2\right)}
\global\long\def\D{\mathcal{D}}
\global\long\def\T{\mathrm{T}}
\global\long\def\Gammafl{\Gamma_{\mathrm{fl}}}
\global\long\def\gammafl{\gamma_{\mathrm{fl}}}
\global\long\def\E#1{\left\langle #1\right\rangle }

\appendix

\section{Derivation of mean-field equation\label{app:Derivation-of-mean-field}}

The generating functional $Z[\bl](\bJ)$ in \prettyref{eq:Z_j} is
properly normalized independent of the realization of $\bJ$. This
property allows us to follow \citep{DeDomincis78_353} and to introduce
the disorder-averaged generating functional
\begin{align}
\bar{Z}[\bl] & :=\langle Z[\bl](\bJ)\rangle_{\bJ}\label{eq:disorder_averaged_Z}\\
 & =\int\Pi_{ij}dJ_{ij}\,\mathcal{N}(0,\frac{g^{2}}{N},J_{ij})\,Z[\bl](\bJ).\nonumber 
\end{align}
The coupling term $\exp\left(-\sum_{i\neq j}J_{ij}\tilde{x}_{i}^{\T}\phi(x_{j})\right)$
in \prettyref{eq:Z_j} factorizes over unit indices $i,j$ and the
random weights $J_{ij}$ appear linear in the exponent. Thus we can
separately integrate over the independently and identically distributed
$J_{ij},\,i\neq j,$ by completing the square and obtain
\begin{eqnarray}
 &  & \int\,dJ_{ij}\mathcal{N}(0,\frac{g^{2}}{N},J_{ij})\,\exp\left(-J_{ij}\tilde{x}_{i}^{\T}\phi(x_{j})\right)\label{eq:completion_of_square}\\
 & = & \exp\left(\frac{g^{2}}{2N}\left(\tilde{x}_{i}^{\T}\phi(x_{j})\right)^{2}\right).\nonumber 
\end{eqnarray}
We reorganize the resulting sum in the exponent of the coupling term
as
\begin{eqnarray}
 &  & \frac{g^{2}}{2N}\sum_{i\neq j}\left(\int\,\tilde{x}_{i}(t)\phi(x_{j}(t))\,dt\right)^{2}\nonumber \\
 & = & \frac{g^{2}}{2N}\sum_{i\neq j}\int\int\,\tilde{x}_{i}(t)\phi(x_{j}(t))\,\tilde{x}_{i}(t^{\prime})\phi(x_{j}(t^{\prime}))\,dt\,dt^{\prime}\nonumber \\
 & = & \frac{1}{2}\sum_{i}\int\int\,\tilde{x}_{i}(t)\tilde{x}_{i}(t^{\prime})\,\left(\frac{g^{2}}{N}\sum_{j}\phi(x_{j}(t))\phi(x_{j}(t^{\prime}))\right)\,dt\,dt^{\prime}\nonumber \\
 &  & -\frac{1}{2}\sum_{i}\int\int\,\tilde{x}_{i}(t)\tilde{x}_{i}(t^{\prime})\,\frac{g^{2}}{N}\phi(x_{i}(t))\phi(x_{i}(t^{\prime}))\,dt\,dt^{\prime},\label{eq:quenched_averaged}
\end{eqnarray}
where we used $\left(\int\,f(t)dt\right)^{2}=\iint\,f(t)f(t^{\prime})\,dt\,dt^{\prime}$
in the first step and $\sum_{ij}x_{i}y_{j}=\sum_{i}x_{i}\sum_{j}y_{j}$
in the second. The last line is the diagonal (self-coupling) to be
skipped in the double sum. It is a correction of order $N^{-1}$ and
will be neglected in the following. The disorder-averaged generating
functional \prettyref{eq:disorder_averaged_Z} therefore takes the
form
\begin{eqnarray}
\bar{Z}[\bl] & = & \int\D\bx\int\D\tbx\,\exp\Big(S_{0}[\bx,\tbx]+\bl^{\T}\bx\Big)\nonumber \\
 &  & \times\exp\Big(\frac{1}{2}\tilde{\bx}^{\T}\,Q_{1}\tilde{\bx}\Big)\,,\label{eq:Zbar_pre}
\end{eqnarray}
where we extended our notation with $x^{\T}Ay:=\iint\,x(t)\,A(t,t^{\prime})\,y(t^{\prime})\,dt\,dt^{\prime}$
to bi-linear forms and defined
\begin{align}
Q_{1}(t,t^{\prime}):= & \frac{g^{2}}{N}\sum_{j}\phi(x_{j}(t))\phi(x_{j}(t^{\prime}))\,.\label{eq:def_Q1}
\end{align}
The field $Q_{1}$ is an empirical average over $N$ contributions,
which, by the law of large numbers and in the case of weak correlations,
will converge to its expectation value for large $N$. This heuristic
argument is shown in the following more formally: A saddle-point approximation
leads to the replacement of $Q_{1}$ by its (self-consistent) expectation
value. To this end we first decouple the interaction term by inserting
the Fourier representation of the Dirac-$\delta$ functional: 
\begin{align}
 & \delta[-\frac{N}{g^{2}}Q_{1}(t,s)+\phi(\bx(t))^{\T}\,\phi(\bx(s))]\label{eq:Hubbard_Stratonovich}\\
= & \int\D Q_{2}\,\exp\left(-\frac{N}{g^{2}}Q_{1}^{\T}Q_{2}+\phi(\bx)^{\T}Q_{2}\phi(\bx)\right),\nonumber 
\end{align}
where we further extended our notation with $Q_{1}^{\T}Q_{2}:=\iint\,Q_{1}(t,s)\,Q_{2}(t,s)\,dt\,ds$
and $\phi(\bx(t))^{\T}\phi(\bx(s))=\sum_{i=1}^{N}\phi(x_{i}(t))\phi(x_{i}(s))$.
We note that the conjugate field $Q_{2}\in i\mathbb{R}$ is purely
imaginary. We hence rewrite \prettyref{eq:Zbar_pre} as\begin{widetext}

\begin{align}
\bar{Z}[j,\tj] & =\int\D Q_{1}\int\D Q_{2}\exp\left(-\frac{N}{g^{2}}Q_{1}^{\T}Q_{2}+N\,\ln\,\Omega[Q_{1},Q_{2}]+j^{\T}Q_{1}+\tj^{\T}Q_{2}\right)\,\label{eq:Zbar}\\
\Omega[Q_{1},Q_{2}] & =\int\D x\int\D\tx\,\exp\Big(S_{0}[x,\tx]+\frac{1}{2}\tx^{\T}Q_{1}\tx+\phi(x)^{\T}Q_{2}\phi(x)\Big),\nonumber 
\end{align}
\end{widetext}where we introduced source terms $j,\tilde{j}$ for
the auxiliary fields and dropped the original source terms $\bl^{\T}x$.
The integral measures $\D Q_{1,2}$ must be defined suitably. In writing
$N\,\ln\,\Omega[Q_{1},Q_{2}]$ we have used that the auxiliary fields
couple only to sums of fields $\sum_{i}\phi^{2}(x_{i})$ and $\sum_{i}\tx_{i}^{2}$,
so that the generating functional for the fields $\bx$ and $\tbx$
factorizes into a product of $N$ identical factors $\Omega[Q_{1},Q_{2}]$. 

The remaining problem can be considered a field theory for the auxiliary
fields $Q_{1}$ and $Q_{2}$. The form \prettyref{eq:Zbar} clearly
exposes the $N$ dependence of the action for these latter fields:
It is of the form $\int dQ\,\exp(Nf(Q))$, which, for large $N$,
suggests a saddle point approximation, which neglects fluctuations
in the auxiliary fields and hence sets them equal to their expectation
value; this point is the dominant contribution to the probability
mass. To obtain the saddle point equations we consider the Legendre-Fenchel
transform of $\ln\,\bar{Z}$ as 
\begin{align*}
\Gamma[q_{1},q_{2}] & :=\sup_{j,\tj}\,j^{\T}q_{1}+\tilde{j}^{\T}q_{2}-\ln\bar{Z}[j,\tj],
\end{align*}
called the vertex generating functional or effective action \citep{ZinnJustin96,NegeleOrland98}.
It holds that $\frac{\delta\Gamma}{\delta q_{1}}=j$ and $\frac{\delta\Gamma}{\delta q_{2}}=\tj$,
called equations of state. The leading order mean-field or tree-level
approximation amounts to the approximation $\Gamma[q_{1},q_{2}]\simeq-S[q_{1},q_{2}]$,
where $S[Q_{1},Q_{2}]=-\frac{N}{g^{2}}Q_{1}^{\T}Q_{2}+N\,\ln\,\Omega[Q_{1},Q_{2}]$
is the action for the auxiliary fields $Q_{1}$ and $Q_{2}$. We insert
this tree-level approximation into the equations of state and further
set $j=\tj=0$ since the source fields have no physical meaning and
thus must vanish. We get the saddle point equations

\begin{align}
0 & =\frac{\delta S[Q_{1},Q_{2}]}{\delta Q_{\{1,2\}}}\label{eq:def_saddle_Q1_Q2}\\
 & =\frac{\delta}{\delta Q_{\{1,2\}}}\left(-\frac{N}{g^{2}}Q_{1}^{\T}Q_{2}+N\,\ln\Omega[Q_{1},Q_{2}]\right)\nonumber 
\end{align}
from which we obtain a pair of equations
\begin{eqnarray}
0 & = & -\frac{N}{g^{2}}\,Q_{1}^{\ast}(t,s)+\frac{N}{\Omega}\,\left.\frac{\delta\Omega[Q_{1},Q_{2}]}{\delta Q_{2}(t,s)}\right|_{Q^{\ast}}\label{eq:saddle_Q1_Q2}\\
\leftrightarrow Q_{1}^{\ast}(t,s) & = & g^{2}\left\langle \phi(x(t))\phi(x(s))\right\rangle _{Q^{\ast}}=:g^{2}C_{\phi(x)\phi(x)}(t,s)\nonumber \\
0 & = & -\frac{N}{g^{2}}\,Q_{2}^{\ast}(t,s)+\frac{N}{\Omega}\,\left.\frac{\delta\Omega[Q_{1},Q_{2}]}{\delta Q_{1}(t,s)}\right|_{Q^{\ast}}\nonumber \\
\leftrightarrow Q_{2}^{\ast}(t,s) & = & \frac{g^{2}}{2}\langle\tilde{x}(t)\tilde{x}(s)\rangle_{Q^{\ast}}=0,\nonumber 
\end{eqnarray}
where we defined the average autocorrelation function $C_{\phi(x)\phi(x)}(t,s)$
of the non-linearly transformed activity of the units. The second
saddle point $Q_{2}^{\ast}=0$ vanishes, because the field was introduced
to represent a Dirac $\delta$ constraint in Fourier domain. One can
show that consequently $\int\D Q\,\exp(S[Q_{1},Q_{2}])Q_{2}=0$, which
is the true mean value $Q_{2}^{\ast}=\langle Q_{2}\rangle=0$, known
as the Deker-Haake theorem.

Here $\langle\rangle_{Q^{\ast}}$ denotes the expectation value with
respect to realizations of $x$ evaluated at the saddle point $Q^{*}$.
The expectation value must be computed self-consistently, since the
values of the saddle points, by \prettyref{eq:Zbar}, influence the
statistics of the fields $\bx$, which in turn determines the function
$Q_{1}^{\ast}$ by \eqref{eq:saddle_Q1_Q2}. Inserting the saddle
point solution into the generating functional \eqref{eq:Zbar} we
get \prettyref{eq:Z_bar_star}
\begin{align*}
\bar{Z}^{\ast} & \propto\int\D x\int\D\tilde{x}\,\exp\,\Big(S_{0}[x,\tx]+\frac{g^{2}}{2}\tx^{\T}C_{\phi(x)\phi(x)}\tx\Big).
\end{align*}
The action has the important property that it decomposes into a sum
of actions for individual, non-interacting units that each feel a
field with a common, self-consistently determined statistics, characterized
by its second cumulant $C_{\phi(x)\phi(x)}$. Prior to the saddle
point approximation \prettyref{eq:Zbar} the fluctuations in the field
$Q_{1}$ are common to all the single units, which effectively couples
them. The saddle-point approximation replaces the fluctuating field
$Q_{1}$ by its mean \prettyref{eq:saddle_Q1_Q2}, which reduces the
network to $N$ non-interacting units, or, equivalently, a single
unit system. The second term in \prettyref{eq:Z_bar_star} is a Gaussian
noise with a two point correlation function $C_{\phi(x)\phi(x)}(t,s)$.
The physical interpretation is the noisy signal each unit receives
due to the input from the other $N$ units. Its autocorrelation function
is given by the summed autocorrelation functions of the output activities
$\phi(x_{i}(t))$ weighted by $g^{2}N^{-1}$, which incorporates the
Gaussian statistics of the couplings. 

The interpretation of the noise can be appreciated by explicitly considering
the moment generating functional of a Gaussian noise with a given
autocorrelation function $C(t,s)$, which leads to the cumulant generating
functional $\ln Z_{\eta}[\tilde{x}]$ that appears in the exponent
of \prettyref{eq:Z_bar_star} and has the form
\begin{eqnarray*}
\ln\,Z_{\eta}[-\tilde{x}] & = & \ln\langle\exp\left(-\tx^{T}\eta\right)\rangle\\
 & = & \frac{1}{2}\tx^{\T}\,C\,\tx.
\end{eqnarray*}
Note that the only non-vanishing cumulant of the effective noise is
the second cumulant; the cumulant generating functional is quadratic
in $\tx$. This means the effective noise is Gaussian and only couples
pairs of time points in proportion to the correlation function.

\section{Stationary process\label{app:Stationary-process}}

We rewrite equation \prettyref{eq:mean-field_diffeq} as 

\begin{eqnarray}
(\partial_{t}+1)\,x(t) & = & \tilde{\eta}(t),\label{eq:mean-field_diffwq_tilde}
\end{eqnarray}
where we combined the two independent Gaussian processes $\eta$ and
$\xi$ appearing in \prettyref{eq:mean-field_diffeq} into $\tilde{\eta}(t)$.
We then multiply \prettyref{eq:mean-field_diffwq_tilde} for time
points $t$ and $s$ and take the expectation value over realizations
of the noise $\tilde{\eta}$ on both sides, which leads to
\begin{align}
\left(\partial_{t}+1\right)\left(\partial_{s}+1\right)\,C_{xx}(t,s) & =g^{2}\,C_{\phi(x)\phi(x)}(t,s)+2\sigma^{2}\delta(t-s),\label{eq:cov_xx_diffeq}
\end{align}
where we defined the covariance function of the activities $C_{xx}(t,s):=\langle x(t)x(s)\rangle$.
We are now interested in the stationary statistics $C_{xx}(t,s)=:c(t-s)$
of the system. The inhomogeneity in \prettyref{eq:cov_xx_diffeq}
is then also time-translation invariant; $C_{\phi(x)\phi(x)}(t+\tau,t)$
is only a function of $\tau$. Therefore the differential operator
$\left(\partial_{t}+1\right)\left(\partial_{s}+1\right)c(t-s)$, with
$\tau=t-s$, simplifies to $(-\partial_{\tau}^{2}+1)\,c(\tau)$ so
we get
\begin{eqnarray}
(-\partial_{\tau}^{2}+1)\,c(\tau) & = & g^{2}\,C_{\phi(x)\phi(x)}(t+\tau,t)+2\sigma^{2}\,\delta(\tau),\nonumber \\
\label{eq:diffeq_auto}
\end{eqnarray}
given as \prettyref{eq:diffeq_c} in the main text.

\section{Replica calculation to assess the Lyapunov exponent\label{app:pair_of_systems}}

We start from the generating functional for the pair of systems \prettyref{eq:Z_pair}
and perform the average over realizations of the connectivity $\bJ$,
as in \prettyref{eq:completion_of_square}. We therefore need to evaluate
the Gaussian integral
\begin{align}
 & \int dJ_{ij}\mathcal{N}(0,\frac{g^{2}}{N},J_{ij})\,\exp\left(-J_{ij}\sum_{\alpha=1}^{2}\tilde{x}_{i}^{\alpha\T}\phi(x_{j}^{\alpha})\right)\nonumber \\
 & =\exp\left(\frac{g^{2}}{2N}\sum_{\alpha=1}^{2}\big[\tilde{x}_{i}^{\alpha\T}\phi(x_{j}^{\alpha})\big]^{2}\right)\nonumber \\
 & \times\exp\left(\frac{g^{2}}{N}\,\big[\tilde{x}_{i}^{1\T}\phi(x_{j}^{1})\big]\,\big[\tilde{x}_{i}^{2\T}\phi(x_{j}^{2})\big]\right).\label{eq:quenched_avg_pair}
\end{align}
The first exponential factor only includes variables of a single subsystem
and is identical to the term appearing in \prettyref{eq:Zbar_pre}.
The second exponential factor is a coupling term between the two systems
arising from the identical matrix $\bJ$ in the two replicas in each
realization that enters the expectation value. We treat the former
terms as before and here concentrate on the mixed coupling term. Analogous
to \prettyref{eq:quenched_averaged}, the exponent of the mixed coupling
term can be rewritten as
\begin{align}
 & \frac{g^{2}}{N}\sum_{i\neq j}\,\big[\tilde{x}_{i}^{1\T}\phi(x_{j}^{1})\big]\,\big[\tilde{x}_{i}^{2\T}\phi(x_{j}^{2})\big]\label{eq:mixed_avg_pair}\\
= & \iint\,\sum_{i}\tilde{x}_{i}^{1}(t)\tilde{x}_{i}^{2}(s)\frac{g^{2}}{N}\sum_{j}\phi(x_{j}^{1}(t))\,\phi(x_{j}^{2}(s))\,dt\,ds+O(N^{-1}),\nonumber 
\end{align}
where we included the self coupling term $i=j$, which is again a
subleading correction of order $N^{-1}$.

We now follow the steps in \prettyref{app:Derivation-of-mean-field}
and introduce three pairs of auxiliary variables. The pairs $Q_{1}^{\alpha},Q_{2}^{\alpha}$
are defined as before in \prettyref{eq:def_Q1} and \prettyref{eq:Hubbard_Stratonovich},
but for each subsystem, while the pair $T_{1},T_{2}$ decouples the
mixed term \prettyref{eq:mixed_avg_pair} by defining
\begin{align*}
T_{1}(t,s) & :=\frac{g^{2}}{N}\sum_{j}\phi(x_{j}^{1}(t))\,\phi(x_{j}^{2}(s)).
\end{align*}
Taken together, we can therefore rewrite the generating functional
\prettyref{eq:Z_pair} averaged over the couplings as\begin{widetext}
\begin{align}
\bar{Z}:=\langle Z(\bJ)\rangle_{\bJ} & =\Pi_{\alpha=1}^{2}\left\{ \int\D Q_{1}^{\alpha}\int\D Q_{2}^{\alpha}\right\} \int\D T_{1}\int\D T_{2}\,\exp\Big(S[\{Q_{1}^{\alpha},Q_{2}^{\alpha}\}_{\alpha\in\{1,2\}},T_{1},T_{2}]\Big)\label{eq:Zbar_pair_HS}\\
S[\{Q_{1}^{\alpha},Q_{2}^{\alpha}\}_{\alpha\in\{1,2\}},T_{1},T_{2}] & :=-\sum_{\alpha=1}^{2}\frac{N}{g^{2}}Q_{1}^{\alpha\T}Q_{2}^{\alpha}-\frac{N}{g^{2}}T_{1}^{\T}T_{2}+N\,\ln\,\Omega^{12}[\{Q_{1}^{\alpha},Q_{2}^{\alpha}\}_{\alpha\in\{1,2\}},T_{1},T_{2}]\nonumber \\
\Omega^{12}[\{Q_{1}^{\alpha},Q_{2}^{\alpha}\}_{\alpha\in\{1,2\}},T_{1},T_{2}] & =\Pi_{\alpha=1}^{2}\Big\{\int\D x^{\alpha}\int\D\tilde{x}^{\alpha}\,\exp\Big(S_{0}[x^{\alpha},\tilde{x}^{\alpha}]+\frac{1}{2}\tilde{x}^{\alpha\T}Q_{1}^{\alpha}\tilde{x}^{\alpha}+\phi(x{}^{\alpha})^{\T}Q_{2}^{\alpha}\phi(x{}^{\alpha})\Big)\Big\}\nonumber \\
 & \times\exp\left(\tilde{x}^{1\T}\left(T_{1}+2\sigma^{2}\right)\tilde{x}^{2}+\phi(x^{1})^{\T}T_{2}\phi(x^{2})\Big)\right)\,,\nonumber 
\end{align}
\end{widetext}where we used that the generating functional factorizes
into a product of $2N$ identical factors $Z^{12}$.

Analogously to \prettyref{app:Derivation-of-mean-field} we could
introduce sources for the auxiliary fields $Q_{1}^{\alpha}$, $Q_{2}^{\alpha}$,
$T_{1}$, $T_{2}$. Then the equations of state are obtained from
the vertex-generating functional $\Gamma$ as before, which, in the
tree-level approximation is given by $\Gamma=-S$ and for vanishing
sources leads to the saddle-point equations $\frac{\delta S}{\delta Q_{1,2}^{\alpha}}=\frac{\delta S}{\delta T_{1,2}}\stackrel{!}{=}0$.
From the latter we obtain the set of equations

\begin{align}
Q_{1}^{\alpha\ast}(t,s) & =g^{2}\,\frac{1}{\Omega^{12}}\,\frac{\delta\Omega^{12}}{\delta Q_{2}^{\alpha}(t,s)}=g^{2}\,\langle\phi(x^{\alpha}(t))\phi(x^{\alpha}(s))\rangle_{Q^{\ast},T^{\ast}}\label{eq:saddle_pair}\\
Q_{2}^{\alpha\ast}(t,s) & =0\nonumber \\
T_{1}^{\ast}(t,s) & =g^{2}\,\frac{1}{\Omega^{12}}\,\frac{\delta\Omega^{12}}{\delta T_{2}(t,s)}=g^{2}\,\langle\phi(x^{1}(t))\phi(x^{2}(s))\rangle_{Q^{\ast},T^{\ast}}\nonumber \\
T_{2}^{\ast}(t,s) & =0.\nonumber 
\end{align}
The generating functional at the saddle point therefore is
\begin{align}
\bar{Z}^{\ast} & =\iint\,\Pi_{\alpha=1}^{2}\D x^{\alpha}\D\tilde{x}^{\alpha}\exp\Big(\sum_{\alpha=1}^{2}S_{0}[x^{\alpha},\tilde{x}^{\alpha}]+\frac{1}{2}\tilde{x}^{\alpha\T}Q_{1}^{\alpha\ast}\tilde{x}^{\alpha}\Big)\nonumber \\
 & \times\exp\left(\tilde{x}^{1\T}\left(T_{1}^{\ast}+2\sigma^{2}\right)\tilde{x}^{2}\right).\label{eq:Z_bar_pair_ast}
\end{align}
We make the following observations: 
\begin{enumerate}
\item The two subsystems $\alpha=1,2$ in the first line of \prettyref{eq:Z_bar_pair_ast}
have the same form as in \prettyref{eq:Z_bar_star}. This has been
expected, because there is no physical coupling between the two systems.
This implies that the marginal statistics of the activity in one system
cannot be affected by the mere presence of the second. Hence in particular
the saddle points $Q_{1,2}^{\alpha\ast}$ must be the same as in \prettyref{eq:Z_bar_star}.
\item If the term in the second line of \eqref{eq:Z_bar_pair_ast} was absent,
the statistics in the two systems would be independent. Two sources,
however, contribute to correlations between the systems: The common
Gaussian white noise that gives rise to the term $\propto2\sigma^{2}$
and the non-white Gaussian noise due to a non-zero value of the auxiliary
field $T_{1}^{\ast}(t,s)$.
\item Only products of pairs of fields appear in \prettyref{eq:Z_bar_pair_ast},
so that the statistics of the $x^{\alpha}$ is Gaussian.
\end{enumerate}
From \prettyref{eq:Z_bar_pair_ast} and \prettyref{eq:saddle_pair}
we can read off the pair of effective dynamical equations \prettyref{eq:effective_pair_eq}
with self-consistent statistics \prettyref{eq:Leffective_pair_noise}.

\subsection{Derivation of the variational equation\label{app:Derivation-of-variational}}

We multiply the equation \prettyref{eq:effective_pair_eq} for $\alpha=1$
and $\alpha=2$ and take the expectation value on both sides, so we
get for $\alpha,\beta\in\{1,2\}$
\begin{align}
\left(\partial_{t}+1\right)\left(\partial_{s}+1\right)c^{\alpha\beta}(t,s) & =\nonumber \\
2\sigma^{2}\delta(t-s)+g^{2}F_{\phi}\left(c^{\alpha\beta}(t,s),c^{\alpha\alpha}(t,t),c^{\beta\beta}(s,s)\right)\, & ,\label{eq:diffeq_cab}
\end{align}
where the function $F_{\phi}$ is defined as the expectation value
\begin{align*}
F_{\phi}(c^{12},c^{11},c^{22}) & :=\E{\phi(x^{1})\phi(x^{2})}
\end{align*}
for the centered bi-variate Gaussian distribution
\begin{align*}
\begin{pmatrix}x^{1}\\
x^{2}
\end{pmatrix} & \sim\mathcal{N}_{2}\left(0,\begin{pmatrix}c^{11} & c^{12}\\
c^{12} & c^{22}
\end{pmatrix}\right).
\end{align*}
First, we observe that the equations for the autocorrelation functions
$c^{\alpha\alpha}(t,s)$ decouple and can each be solved separately,
leading to the same equation \prettyref{eq:equation_of_motion} as
before. This formal result could have been anticipated, because the
marginal statistics of each subsystem cannot be affected by the mere
presence of the respective other system. Their solutions
\begin{align*}
c^{11}(t,s)= & c^{22}(t,s)=c(t-s)
\end{align*}
then provide the ``background'' for the equation for the cross-correlation
function between the two copies; they fix the second and third argument
of the function $F_{\phi}$ on the right-hand side of \prettyref{eq:diffeq_cab}.
It remains to determine the equation of motion for $c^{12}(t,s)$.

We first determine the stationary solution $c^{12}(t,s)=k(t-s)$.
We see immediately from \prettyref{eq:diffeq_cab} that $k(\tau)$
obeys the same equation of motion as $c(\tau)$, so $k(\tau)=c(\tau)$
is a solution. The distance \prettyref{eq:mean-squared-distance}
between replicas for this solution therefore vanishes; the dynamics
in both replicas follows identical realizations. Let us now study
the stability of this solution. We hence need to expand $c^{12}$
around the stationary solution
\begin{align*}
c^{12}(t,s) & =c(t-s)+\epsilon\,k^{(1)}(t,s)\,,\:\epsilon\ll1\,.
\end{align*}
We expand the right hand side of \prettyref{eq:diffeq_cab} into a
Taylor series using Price's theorem and \prettyref{eq:def_f} 
\begin{align*}
F_{\phi}\left(c^{12}(t,s),c_{0},c_{0}\right) & =f_{\phi}\left(c^{12}(t,s),c_{0}\right)\\
 & =f_{\phi}\left(c(t-s),c_{0}\right)\\
 & +\epsilon\,f_{\phi^{\prime}}\left(c(t-s),c_{0}\right)\,k^{(1)}(t,s)+O(\epsilon^{2})\,.
\end{align*}
Inserted into \prettyref{eq:diffeq_cab} and using that $c$ solves
the lowest order equation, we get the linear equation of motion for
the first order deflection \prettyref{eq:variational_equation}. By
\prettyref{eq:mean-squared-distance} the first order deflection $k^{(1)}(t,s)$
determines the distance between the two subsystems as 

\begin{align}
d(t) & =\underbrace{c^{11}(t,t)}_{c_{0}}+\underbrace{c^{22}(s,s)}_{c_{0}}\underbrace{-c^{12}(t,t)-c^{21}(t,t)}_{-2c_{0}-2\epsilon\,k^{(1)}(t,t)}\nonumber \\
 & =-2\epsilon\,k^{(1)}(t,t).\label{eq:relation_distance_c_12}
\end{align}
The negative sign makes sense, since we expect in the chaotic state
that $c^{12}(t,s)\stackrel{t,s\to\infty}{=}0$, so $k^{(1)}$ must
be of opposite sign than $c>0$.

\subsection{Schrödinger equation for the maximum Lyapunov exponent\label{app:Schroedinger-equation}}

We here want to reformulate the equation for the variation of the
cross-system correlation \prettyref{eq:variational_equation} into
a Schrödinger equation, as in the original work \citep[eq. 10]{Sompolinsky88_259}.

First, noting that $C_{\phi^{\prime}\phi^{\prime}}(t,s)=f_{\phi^{\prime}}\left(c(t-s),c_{0}\right)$
is time translation invariant, it is advantageous to introduce the
coordinates $T=t+s$ and $\tau=t-s$ and write the covariance $k^{(1)}(t,s)$
as $k(T,\tau)$ with $k^{(1)}(t,s)=k(t+s,t-s)$. The differential
operator $\left(\partial_{t}+1\right)\left(\partial_{s}+1\right)$
with the chain rule $\partial_{t}\to\partial_{T}+\partial_{\tau}$
and $\partial_{s}\to\partial_{T}-\partial_{\tau}$ in the new coordinates
is $(\partial_{T}+1)^{2}-\partial_{\tau}^{2}$. A separation ansatz
$k(T,\tau)=e^{\frac{1}{2}\kappa T}\,\psi(\tau)$ then yields the eigenvalue
equation
\begin{align*}
(\frac{\kappa}{2}+1)^{2}\psi(\tau)-\partial_{\tau}^{2}\psi(\tau) & =g^{2}f_{\phi^{\prime}}\left(c(\tau),c_{0}\right)\psi(\tau)
\end{align*}
for the growth rates $\kappa$ of $d(t)=-2\epsilon k^{(1)}(t,t)=-2\epsilon k(2t,0)$.
We can express the right hand side by the second derivative of the
potential \prettyref{eq:classic_potential} so that with 
\begin{eqnarray}
V^{\prime\prime}(c(\tau);c_{0}) & = & -1+g^{2}f_{\phi^{\prime}}\left(c(\tau),c_{0}\right)\label{eq:effective_potential}
\end{eqnarray}
we get the time-independent Schrödinger equation
\begin{eqnarray}
\left(-\partial_{\tau}^{2}-V^{\prime\prime}(c(\tau);c_{0})\right)\,\psi(\tau) & = & \underbrace{\left(1-\left(\frac{\kappa}{2}+1\right)^{2}\right)}_{=:E}\,\psi(\tau),\nonumber \\
\label{eq:Schroedinger_appendix}
\end{eqnarray}
where the time lag $\tau$ plays the role of a spatial coordinate
for the Schrödinger equation. The eigenvalues (``energies'') $E_{n}$
determine the exponential growth rates $\kappa_{n}$ of the solutions
$k(2t,0)=e^{\kappa_{n}t}\,\psi_{n}(0)$ at $\tau=0$ with 
\begin{eqnarray}
\kappa_{n}^{\pm} & = & 2\left(-1\pm\sqrt{1-E_{n}}\right).\label{eq:roots_lambda}
\end{eqnarray}
We can therefore determine the growth rate of the mean-square distance
of the two subsystems by \prettyref{eq:relation_distance_c_12}. The
fastest growing mode of the distance is hence given by the ground
state energy $E_{0}$ and the plus in \prettyref{eq:roots_lambda}.
The deflection between the two subsystems therefore growth with the
rate
\begin{eqnarray}
\lambda_{\mathrm{max}} & = & \frac{1}{2}\kappa_{0}^{+}\label{eq:Lambda_max}\\
 & = & -1+\sqrt{1-E_{0}},\nonumber 
\end{eqnarray}
where the factor $1/2$ in the first line is due to $d$ being the
squared distance, hence the length $\sqrt{d}$ growth with half the
exponent as $d$.

\section{Memory curve\label{app:Memory-curve}}

To evaluate \prettyref{eq:memory_def} we need to determine the disorder-averaged
sum of squared response functions

\begin{equation}
\sum_{i=1}^{K}\overline{\E{x_{i}(t)z(t_{0})}^{2}}\,,\label{eq:memory}
\end{equation}
with $t_{0}=t-\tau$ and $K$ denoting the number of neurons connected
to the readout, which we initially leave as a free parameter. Here
$\langle\rangle$ denotes the average over realizations of the inputs
$\xi_{i}$ (or alternatively over time) and the overbar the average
over realizations of the connectivity $\bJ$ as in \prettyref{eq:disorder_averaged_Z}.
Moreover, we here examine a more general input signal $z(t)=\sum_{j=1}^{N}v_{j}\xi_{j}(t)$
where $\boldsymbol{v}$ denotes the input weights.

We pick two points in time $t,s\ge t_{0}$ and define
\begin{equation}
h^{K}(t,s):=\sum_{i=1}^{K}\overline{\E{x_{i}(t)z(t_{0})}\E{x_{i}(s)z(t_{0})}}\,.\label{eq:def_hk}
\end{equation}
The measure of interest, \prettyref{eq:memory}, then follows for
$t=s$. The key idea is to express the correlator $\E{x_{i}(t)z(t_{0})}$
as a weighted sum of response functions $\langle x_{i}(t)\tx(t_{0})\rangle$,
which we show in the following. We introduce a scalar source term
$k(t)$ for the signal $z(t)$ and average over the noise $\xi$.
This yields the generating functional
\begin{align}
Z[l,k] & =\int\D\boldsymbol{x}\,\int\D\boldsymbol{\tilde{x}}\,\times\label{eq:Z_single-1}\\
 & \times\exp(\sum_{i}\,S_{0}[x_{i},\tilde{x}_{i}]-\sum_{j}J_{ij}\,\phi(x_{j})-2\sigma^{2}v_{i}\,k^{\T}\tx_{i}+l_{i}^{\T}x_{i}).\nonumber 
\end{align}
Evaluating the correlator leads to 
\begin{align}
\langle x_{i}(t)z(t_{0})\rangle & =\frac{\delta^{2}Z}{\delta l_{i}(t)\delta k(t_{0})}\Big|_{j_{i}=k=0}\label{eq:xz_as_response}\\
 & =-2\sigma^{2}\sum_{j=1}^{N}v_{j}\langle x_{i}(t)\tx_{j}(t_{0})\rangle\,.\nonumber 
\end{align}
We now consider a pair of systems (replicas) similarly as in \prettyref{app:pair_of_systems}
with the difference, however, that the two systems receive independent
realizations of the inputs $\xi$. We need two independent systems
to express the product of the two correlators in \prettyref{eq:memory}.
By independence, the corresponding average factorizes, 
\begin{eqnarray}
\E{x_{i}(t)z(t_{0})}\E{x_{i}(s)z(t_{0})} & = & \E{x_{i}^{1}(t)z^{1}(t_{0})\,x_{i}^{2}(s)z^{2}(t_{0})}\nonumber \\
 & =: & h_{i}(t,s,t_{0}),\label{eq:four_point_correlator}
\end{eqnarray}
where the superscript denotes the replicon index as before. To obtain
$h$, it is sufficient to introduce a single source term
\begin{align}
 & 4\sigma^{4}\sum_{j,l}v_{j}v_{l}\int dt\epsilon(t)\tx_{j}^{1}(t)\tx_{l}^{2}(t)\label{eq:gen_func_perturbed}
\end{align}
with source $\epsilon(t)$ to the corresponding generating functional,
which allows us to obtain $\langle x_{i}(t)z(t_{0})\rangle$, with
\eqref{eq:xz_as_response} and \eqref{eq:four_point_correlator} as
\begin{align*}
\overline{\E{x_{i}(t)z(t_{0})}\E{x_{i}(s)z(t_{0})}} & =\frac{\delta}{\delta l_{i}^{1}(t)}\frac{\delta}{\delta l_{i}^{2}(s)}\frac{\delta}{\delta\epsilon(t_{0})}Z\big|_{l=\epsilon=0}.
\end{align*}
The additional source term \eqref{eq:gen_func_perturbed} has the
physical interpretation of a common input with time-dependent variance
$\epsilon(t)$ injected into a pair of units between the two replicas.
The absence of quadratic terms $\propto(\tx^{\alpha})^{2}$ shows
that this common input does not affect the marginal statistics of
the two systems in isolation. This interpretation is here only mentioned
for illustrative purposes; the derivation does not rely on it.Due
to the weight $v_{j}v_{l}$ for different unit pairs $j,l$ we keep
the single neuron index in the following.

The goal now is to derive a differential equation for the disorder
averaged $\overline{h_{i}(t,s,t_{0})}$ similar to \prettyref{app:Derivation-of-variational},
needed to compute \prettyref{eq:memory}.

First, after averaging over the disorder, completely analogous to
\prettyref{app:pair_of_systems}, we can read off effective equations
for the single units

\begin{align}
\left(\partial_{t}+1\right)x_{i}^{\alpha}(t) & =\xi_{i}^{\alpha}(t)+\eta_{i}^{\alpha}(t)+\rho_{i}^{\alpha}(t)\,\label{eq:eff_diffeq_mem_ap}
\end{align}
$\alpha\in\{1,2\},\,i\in\{1,...,N\},$ together with a set of self-consistency
equations for the statistics of the noises
\begin{align}
\langle\xi_{i}^{\alpha}(t)\,\xi_{j}^{\beta}(s)\rangle & =2\sigma^{2}\delta_{\alpha\beta}\delta_{ij}\delta(t-s)\nonumber \\
\langle\eta_{i}^{\alpha}(t)\,\eta_{j}^{\beta}(s)\rangle & =\frac{g^{2}}{N}\delta_{ij}\,\sum_{i}\langle\phi(x_{i}^{\alpha}(t))\phi(x_{i}^{\beta}(s))\rangle\nonumber \\
\langle\rho_{i}^{\alpha}(t)\,\rho_{j}^{\beta}(s)\rangle & =4\sigma^{4}(1-\delta_{\alpha\beta})\,\nu_{i}\nu_{j}\epsilon(t)\,\delta(t-s)\,.\label{eq:noise_stat}
\end{align}
The first line in \prettyref{eq:noise_stat} represents the independent
noise between the systems, the second line the common connectivity
and the third line the common noise component we introduced in \prettyref{eq:gen_func_perturbed}
to express the squared response function \prettyref{eq:def_hk}.

Second, we obtain $h_{i}(t,s,t_{0})$ by a functional derivative with
respect to $l_{i}^{1}(t),\,l_{i}^{2}(s)$ and $\epsilon(t_{0})$,
which can be seen from its representation as the four-point correlator
in \prettyref{eq:four_point_correlator}. Writing the functional derivative
with respect to $\epsilon$ explicitly as a limit, we can express
$h$ by the correlation between the pair of systems

\begin{align}
\overline{h_{i}(t,s,t_{0})} & =\lim_{\iota\rightarrow0}\frac{1}{\iota}\langle x_{i}^{1}(t)x_{i}^{2}(s)\rangle\Big|_{\epsilon=\iota\,\delta(\circ-t_{0})]}\,,\label{eq:hi_corr_ap}
\end{align}
where we used that for $\epsilon=0$ the two systems are uncorrelated.
We now combine the effective equation \prettyref{eq:eff_diffeq_mem_ap}
and \prettyref{eq:hi_corr_ap} to obtain a partial differential equation
for $h\,$:
\begin{align}
 & \left(\partial_{t}+1\right)\left(\partial_{s}+1\right)\,\overline{h_{i}(t,s,t_{0})}\label{eq:pde_mem_ap}\\
= & \frac{g^{2}}{N}\,\lim_{\iota\rightarrow0}\frac{1}{\iota}\Big[\sum_{i=1}^{N}\langle\phi(x_{i}^{1}(t))\phi(x_{i}^{2}(s))\rangle\Big]\Big|_{\epsilon=\iota\,\delta(\circ-t_{0})}\nonumber \\
+ & 4\sigma^{4}\nu_{i}^{2}\delta(t-s)\,\delta(t-t_{0}).\nonumber 
\end{align}
Since we are interested in the limit $\iota\to0$, we expand the first
term to linear order around the uncorrelated state

\begin{align}
\langle\phi(x_{i}^{1}(t))\phi(x_{i}^{2}(s))\rangle & =f_{\phi}\left(0,c_{0}\right)+\partial_{1}f_{\phi}\left(0,c_{0}\right)c_{i}^{12}(t,s)\nonumber \\
 & =\overline{\langle\phi^{\prime}(x_{i})\rangle}^{2}\,\overline{h_{i}(t,s,t_{0})}\,,\label{eq:linearize_c12}
\end{align}
where the first term vanishes as it factorizes into $\overline{\langle\phi(x)\rangle}^{2}=0$.
Inserting \prettyref{eq:linearize_c12} into \prettyref{eq:pde_mem_ap}
we arrive at

\begin{eqnarray*}
\left(\partial_{t}+1\right)\left(\partial_{s}+1\right)\,\overline{h_{i}(t,s,t_{0})} & = & \sum_{i}\overline{\langle\phi^{\prime}(x_{i})\rangle}^{2}\,\overline{h_{i}(t,s,t_{0})}\\
 & + & 4\sigma^{4}\,\nu_{i}^{2}\,\delta(t-t_{0})\,\delta(s-t_{0})\\
 & = & \overline{\langle\phi^{\prime}(x_{j})\rangle}^{2}\sum_{i}^{N}\overline{h_{i}(t,s,t_{0})}\\
 & + & 4\sigma^{4}\,\nu_{i}^{2}\,\delta(t-t_{0})\,\delta(s-t_{0})\,.
\end{eqnarray*}
In the latter step we used that $\overline{\langle\phi^{\prime}(x_{i})\rangle}$
is independent of $i$ because the expectation value is taken with
respect to the disorder averaged unperturbed system and thus we use
a representative unit $j$ as the index. Taking the sum with respect
to $i=1,...,K$ yields

\begin{align}
\left(\partial_{t}+1\right)\left(\partial_{s}+1\right)\,h^{K}(t,s) & =g^{2}\overline{\E{\phi'\left(x_{j}\right)}}^{2}\frac{K}{N}\,h^{N}(t,s)\label{eq:memory-pde-partial}\\
 & +4\sigma^{4}\|\boldsymbol{v}_{K}\|^{2}\delta(t-t_{0})\delta(s-t_{0})\nonumber 
\end{align}
with $\|\boldsymbol{v}_{K}\|^{2}=\sum_{i=1}^{K}v_{i}^{2}\,$. For
the complete sum of squared response functions, $h^{N}(t,s)$, the
following closed linear partial differential equation holds: 
\begin{align}
\left(\partial_{t}+1\right)\left(\partial_{s}+1\right)h^{N}(t,s) & =g^{2}\overline{\E{\phi'\left(x_{j}\right)}}^{2}h^{N}(t,s)\label{eq:memory-pde-complete}\\
 & +4\sigma^{4}\delta(t-t_{0})\delta(s-t_{0})\,,\nonumber 
\end{align}
where we set $\|\boldsymbol{v}_{N}\|^{2}=\|\boldsymbol{v}\|^{2}=1$
without loss of generality. The solution to this equation describes
the shape of the memory curve if the readout has access to the states
of all neurons. To determine $h^{K}(t,s)$ we note that the difference
$h^{K}(t,s)-\frac{K}{N}h^{N}(t,s)$ is proportional to the solution
of
\begin{align}
\left(\partial_{t}+1\right)\left(\partial_{s}+1\right)\,h^{(0)}(t,s) & =\delta(t-t_{0})\delta(s-t_{0})\,,\label{eq:memory-pde-trivial}
\end{align}
which by direct integration yields
\begin{align}
h^{(0)}(t,s) & =e^{-(t-t_{0})}\Theta(t-t_{0})e^{-(s-t_{0})}\Theta(s-t_{0})\,.\label{eq:memory-shape-trivial}
\end{align}
Thus, $h^{K}(t,s)$ is given by
\begin{align}
h^{K}(t,s) & =\frac{K}{N}h^{N}(t,s)+4\sigma^{4}\left(\|\boldsymbol{v}_{K}\|^{2}-\frac{K}{N}\right)h^{(0)}(t,s)\nonumber \\
 & =4\sigma^{4}\frac{K}{N}h^{(1)}(t,s)+4\sigma^{4}\left(\|\boldsymbol{v}_{K}\|^{2}-\frac{K}{N}\right)h^{(0)}(t,s)\,,\label{eq:memorv-shape-combined}
\end{align}
where $h^{(1)}(t,s)$ solves
\begin{align}
\left(\partial_{t}+1\right)\left(\partial_{s}+1\right)h^{(1)}(t,s) & =a^{2}h^{(1)}(t,s)+\delta(t-t_{0})\delta(s-t_{0})\label{eq:memory-pde}
\end{align}
with parameter $a^{2}=g^{2}\overline{\E{\phi'(x_{j})}}^{2}=1-1/\tau_{\infty}^{2}\,$.
Here $\tau_{\infty}$ is the time scale of the asymptotic decay of
the autocorrelation function.

As in \prettyref{app:Schroedinger-equation} it is useful to change
coordinates to $T=t+s-2t_{0}$ and $\tau=t-s$. In these coordinates
\prettyref{eq:memory-pde} takes the form
\begin{align*}
(\partial_{T}+1)^{2}h^{(1)}(T,\tau)-\partial_{\tau}^{2} & h^{(1)}(T,\tau)=a^{2}h^{(1)}(T,\tau)+2\delta(T)\delta(\tau)
\end{align*}
and setting $h^{(1)}(T,\tau)=e^{-T}u(T,\tau)$ simplifies the PDE
further to
\begin{align}
\partial_{T}^{2}u(T,\tau)-\partial_{\tau}^{2} & u(T,\tau)=a^{2}u(T,\tau)+2\delta(T)\delta(\tau)\,,\label{eq:klein-gordon-1}
\end{align}
a Klein-Gordon wave equation with temporal coordinate $T$ and spatial
coordinate $\tau$ (and negative squared mass $-a^{2}$). We are looking
for the solution $u(T,\tau)$ in $T\ge0,\tau\in\mathbb{R}$. To this
end we consider the temporal Laplace and the spatial Fourier transform
of \prettyref{eq:klein-gordon-1}. Fourier transformation in $\tau$
yields
\begin{align}
\partial_{T}^{2}\hat{u}(T,k)+\left(k^{2}-a^{2}\right) & \hat{u}(T,k)=2\delta(T)\label{eq:klein-gordon-ft-1}
\end{align}
with $k\in\mathbb{R}$ and the Fourier representation
\begin{align*}
u(T,\tau) & =\frac{1}{2\pi}\intop_{-\infty}^{+\infty}e^{ik\tau}\hat{u}(T,k)\,dk\,.
\end{align*}
For each $k\in\mathbb{R}$ the Laplace transformation in $T$,
\begin{align*}
\tilde{u}(p,k) & =\intop_{0}^{\infty}e^{-pT}\hat{u}(T,k)\,dT\,,
\end{align*}
of \prettyref{eq:klein-gordon-ft-1} reads
\begin{align*}
p^{2}\tilde{u}(p,k)-p\underbrace{\hat{u}(0,k)}_{=0}-\underbrace{\partial_{T}\hat{u}(0,k)}_{=0}+\left(k^{2}-a^{2}\right) & \tilde{u}(p,k)=2\,.
\end{align*}
Hence, in the Fourier-Laplace domain we obtain
\begin{align*}
\tilde{u}(p,k) & =\frac{2}{p^{2}+k^{2}-a^{2}}\,.
\end{align*}
For the memory curve we only need the solution $u(T,\tau)$ for $\tau=0$,
the diagonal $s=t$ in the original coordinates. Setting $\tau=0$
in the Fourier representation gives the Laplace transform of $u(T,\tau=0)$:
\begin{align}
\tilde{u}(p)=\tilde{u}(p,\tau=0) & =\frac{1}{\pi}\intop_{-\infty}^{+\infty}\frac{1}{k^{2}+p^{2}-a^{2}}\,dk\nonumber \\
 & =\frac{1}{\sqrt{p^{2}-a^{2}}}\label{eq:kg-soln-diag-laplace-1}
\end{align}
with $p\in\mathbb{C}$ such that $\mathrm{Re}\left(p^{2}\right)>a^{2}$.
The function on the right is the Laplace transform of the modified
Bessel function of the first kind $I_{0}(aT)$ \citep{Abramowitz74}.
Together with $h^{(1)}(T,\tau)=e^{-T}u(T,\tau)$ we therefore obtain
the shape of the memory curve as
\begin{align}
h^{(1)}(T)=h^{(1)}(T,\tau=0) & =e^{-T}\,I_{0}(aT)\,\Theta(T)\,.\label{eq:memory-shape-complete}
\end{align}

Finally, using \prettyref{eq:memory-shape-trivial} and \prettyref{eq:memory-shape-complete}
in \prettyref{eq:memorv-shape-combined} gives the following explicit
expression (setting $t_{0}=0$) for the sum of squared response functions
\begin{align}
h_{K}(t)=h_{K}(t,t) & =4\sigma^{4}\frac{K}{N}e^{-2t}\,I_{0}(a2t)\,\Theta(t)\nonumber \\
 & +4\sigma^{4}\left(\|\boldsymbol{v}_{K}\|^{2}-\frac{K}{N}\right)e^{-2t}\,\Theta(t)\nonumber \\
 & =4\sigma^{4}\frac{K}{N}e^{-2t}\left(I_{0}(a2t)-1\right)\Theta(t)\label{eq:memory_final}\\
 & +4\sigma^{4}\|\boldsymbol{v}_{K}\|^{2}e^{-2t}\,\Theta(t)\,.\nonumber 
\end{align}
In \prettyref{eq:memory_final} we split $h_{K}(t)$ into two contributions:
a network contribution $h_{K}^{\mathrm{net}}(t)$ proportional to
$K/N$ with shape $e^{-2t}\left(I_{0}(a2t)-1\right)\Theta(t)$ and
a local contribution proportional to $\|\boldsymbol{v}_{K}\|^{2}$
with shape $e^{-2t}\,\Theta(t)$. The latter is just the memory of
the signal due to the leaky integration of the single units, while
the former describes the memory due to the collective network dynamics;
only this contribution is affected by the network parameters.

We can evaluate \prettyref{eq:memory_def} using \prettyref{eq:memory_final}
with the choice $v_{i}=1/\sqrt{N}\,\forall i$ which leads to \prettyref{eq:memory_curve}
and to the network memory \prettyref{eq:memory_curve_net}.

\end{document}